\documentclass[acmtog,review=false]{acmart}

\setcopyright{acmcopyright}

\acmJournal{TOG}
\acmVolume{VV}
\acmNumber{N}
\acmYear{XXXX}
\acmArticle{XXX}  
\acmDOI{10.1145/XXXXXXX.YYYYYYY}


\usepackage{multirow}
\renewcommand{\vec}{\mathbf}

\usepackage{subfig}
\usepackage{bm}
\usepackage{xcolor}

\DeclareMathOperator*{\argmin}{argmin}

\begin{document}

\title{Redefining A in RGBA: Towards a Standard for Graphical 3D Printing}

\author{Philipp Urban}
\email{philipp.urban@igd.fraunhofer.de}
\affiliation{Fraunhofer Institute for Computer Graphic Research IGD}
\affiliation{Norwegian University of Science and Technology NTNU}

\author{Tejas Madan Tanksale}
\email{tejas.madan.tanksale@igd.fraunhofer.de}

\author{Alan Brunton}
\email{alan.brunton@igd.fraunhofer.de}

\affiliation{Fraunhofer Institute for Computer Graphic Research IGD}

\author{Bui Minh Vu}
\author{Shigeki Nakauchi}
\affiliation{Department of Computer Science and Engineering, Toyohashi University of Technology}

\begin{abstract}
Advances in multimaterial 3D printing have the potential to reproduce various visual appearance attributes of an object in addition to its shape. Since many existing 3D file formats encode color and translucency by RGBA textures mapped to 3D shapes, RGBA information is particularly important for practical applications. In contrast to color (encoded by RGB), which is specified by the object's reflectance, selected viewing conditions and a standard observer, translucency (encoded by A) is neither linked to any measurable physical nor perceptual quantity. Thus, reproducing translucency encoded by A is open for interpretation. 

In this paper, we propose a rigorous definition for A suitable for use in graphical 3D printing, which is independent of the 3D printing hardware and software, and which links both optical material properties and perceptual uniformity for human observers. By deriving our definition from the absorption and scattering coefficients of virtual homogenous reference materials with an isotropic phase function, we achieve two important properties. First, a simple adjustment of A is possible, which preserves the translucency appearance if an object is rescaled for printing. Second, determining the value of A for a real (potentially non-homogenous) material, can be achieved by minimizing a distance function between light transport measurements of this material and simulated measurements of the reference materials. Such measurements can be conducted by commercial spectrophotometers used in graphic arts. 

Finally, we conduct visual experiments employing the method of constant stimuli, and derive from them an embedding of A into a nearly perceptually-uniform scale of translucency for the reference materials.
\end{abstract}


\maketitle


\newcommand{\etal}{\mbox{\emph{et al.\ }}}
\newcommand{\ie}{\mbox{\emph{i.e.}}}

\newcommand{\atten}{{\sigma_t}}
\newcommand{\absorp}{{\sigma_a}}
\newcommand{\scatt}{{\sigma_s}}
\newcommand{\psychatten}{\ifmmode \bm{\atten} \else $\bm{\atten}$~\fi}
\newcommand{\psychA}{\ifmmode \bm{\phi} \else $\bm{\phi}$~\fi}
\newcommand{\meanFreeP}{\ell_{{f}}}
\newcommand{\opticalDepth}{\tau}
\newcommand{\Ahat}{\ifmmode \hat{A} \else $\hat{A}$~\fi}
\newcommand{\A}{\ifmmode A \else $A$~\fi}
\newcommand{\unit}[1]{~\mbox{#1}}
\newcommand{\cm}{\unit{cm}}
\newcommand{\mm}{\unit{mm}}
\newcommand{\nm}{\unit{nm}}
\newcommand{\set}[1]{\ensuremath{\mathcal{#1}}}
\newcommand{\refmats}{\set{R}}

\newcommand{\todo}[1]{\PackageWarning{}{Unprocessed todo}{{\textbf{TODO} \textbf{#1}}}}

\newcommand{\revision}[1]{\textcolor{black}{#1}}

\section{Introduction}
\label{sec:Introduction}
Advances in 3D printing allow the combination of multiple printing materials with different optical properties into a single object at a very high resolution. This allows the reproduction of not only the object's shape but also its visual appearance attributes such as color~\cite{BruntonArikanUrban2015}, translucency~\cite{HasanFuchsMatusikPfisterRusinkiewicz2010,DongWangPellaciniTongGuo2010} or gloss~\cite{BaarSamadzadeganBrettelUrbanOrtizSegovia2014}.

Many existing 3D file formats encode spatially-resolved information of (albedo) color and opacity of an object as a RGBA texture mapped to its 3D geometry. This information is widely used in rendering, where the RGB color information is typically interpreted as device-independent standard RGB~\cite{SuesstrunkBuckleySwen1999} and \A (also called $\alpha$ channel) as a blending or mixing parameter to produce transparent overlays in image composition assuming additive color mixture~\cite{PorterDuff1984}. Such an interpretation is common for 3D file formats, including the 3MF format, which is being pushed by many industry players as a standard for 3D printing: In the core specification~\cite{3MFSpec} the interpretation of \A is left unspecified, whereas in the materials and properties extension~\cite{3MFSpec_Mat_Prop_Ext} additive blending is specified, which however also stipulates that the first color layer must be opaque. Thus, translucent objects are not possible at all according to this interpretation; it is purely a mechanism for additively mixing colors in a specified order. 

Using an additive color mixture model is simple, computationally efficient and robust for on-screen display, but it has severe shortcomings~\cite{Faul2017}: light is altered by matter subtractively not additively, i.e. many real transparent materials cannot be described by this interpretation. As a result, renderings employing \A as an additive mixture parameter are suitable for illustrative purposes rather than accurately simulating the appearance of real objects.

Nonetheless, it is highly desirable to capture perceived translucency of real objects within a single parameter, foremost because it allows the seamless continued use of existing image and 3D file formats, \revision{it is supported by various existing 3D design and image manipulation tools}, but also because the perceptual dimension of translucency is known to be small~\cite{GkioulekasXiaoZhaoAdelsonZicklerBala2013}. Therefore, in this paper we present a new interpretation of the \A channel in RGBA for use in graphical 3D printing \revision{aiming to encode a significant portion of perceived translucency information}.

For the purpose of reproducing translucent objects by 3D printing, a few properties of \A are desired:
\begin{enumerate}
\item \A must be linked to a measurable quantity. Only then, can \A be assigned to real materials via measurements and print material arrangements can be adjusted to match this quantity.
\item For print reproductions, a perceptually uniform scale for \A is important, allowing the minimization of perceived errors rather than physical ones. The viewing condition for this scale must agree with the RGB conditions to ensure consistency of color and translucency. In color printing, the viewing/illuminating geometry (side-lighting) is specified by the International Color Consortium~\cite{ICC43} and is supported by color and spectrophotometric measurement devices employed in graphic arts~\cite{ISO13655}, which are used to calibrate the printers. This rules out backlighting conditions for specifying the perceptual scale, even though the translucency for materials possessing complex light transport properties may appear different for side- and backlighting conditions~\cite{XiaoWalterGkioulekasZicklerAdelsonBala2014}.
\item If an object made of a translucent material is scaled (most commonly shrunk) for printing, it is desirable that average light transport distances can be adjusted accordingly to preserve perceptual translucency cues. Therefore, \A should be adjustable to the print size in relation to the original object size in an intuitive, predictable and computationally efficient way.
\item For design purposes, the absence of cross-contamination between the chromatic channels (chroma and hue) and \A is important, i.e. that changing the chromatic channels has no effect on perceived translucency and vice versa, for the specified viewing conditions. Predictors for chroma and hue can be computed by transforming standard RGB color spaces to the CIELCh color space, CIECAM02~\cite{CIECAM022004} or LAB2000HL~\cite{LissnerUrban2012}.    
\end{enumerate}

Given these observations and constraints, this paper makes the following two main contributions: 
\begin{enumerate}
\item A definition of \A based on a set of virtual reference materials and a measurement methodology for extracting BSSRDF information impacting perceptual translucency cues from real and simulated measurements. This definition allows adjusting \A to the print size so that the print has perceptually similar translucency as the original model.
\item Determining a psychometric function allows us to adjust the definition of \A to be nearly perceptually uniform.
\end{enumerate}
By contrast, this paper does not propose a 3D printing pipeline for fabricating translucent objects. Our goal is to provide a definition that is both physically and perceptually meaningful, compatible with existing data formats, practically measurable for both real materials and 3D printed characterization targets, while remaining as neutral as possible concerning 3D printing hardware and software, thus making it applicable as a \emph{device independent} standard for 3D printing systems capable of reproducing varying degrees of translucency. To illustrate the practicality and expressiveness of our \A definition, in Section \ref{sec:Resizing} and Appendix \ref{sec:additional_prints} we show some 3D printed examples generated using a recently proposed color and translucency 3D printing pipeline~\cite{BruntonArikanTanksaleUrban2018}.

\subsection{Background}
\label{sec:intro:background}

Light reflectance and scattering properties of a non-fluorescent homogenous material are described by the 8-dimensional \emph{Bidirectional Surface Scattering Reflectance Distribution Function} (BSSRDF). It is the ratio of outgoing radiance and incident flux computed for any bidirectional geometry of incoming and outgoing light rays. There exist candidate representations for graphical 3D printing based on approximations or simplifications of the full BSSRDF by factoring the complete function into distinct physical phenomena. Jensen \etal presented a 10-parameter model~\shortcite{JensenMarschnerLevoyHanrahan2001}. The SubEdit model~\cite{song_subedit_scattering_2009} uses 1D scattering profiles to approximate diffuse scattering, and allows interactive rendering and editing of subsurface scattering; examples were shown with a 24-dimensional variant. This representation has even been used in fabrication pipelines~\cite{DongWangPellaciniTongGuo2010,HasanFuchsMatusikPfisterRusinkiewicz2010}. Spatially-varying versions of these representations require either high-dimensional texture or multiple textures mapped onto the surface geometry. More importantly, neither considers human perception of translucency as is done for color, but rather approximate the underlying physical phenomena. 

A BSSRDF allows us to compute stimuli yielding CIEXYZ and standard RGB color space values~\cite{SuesstrunkBuckleySwen1999} for any measurement geometry. In printing, a circular 45/0 measurement geometry is used and supported by many spectrophotometers~\cite{ISO13655}, i.e. the surface is circularly illuminated under $45^{\circ}$ to its normal and the stimulus is measured at $0^{\circ}$. Thus, a material's RGB values contain only a small portion of the BSSRDF information selected already by the measurement geometry.

Analogously, our strategy to define \A is to identify relevant BSSRDF information causing a material to appear more or less translucent. Fleming and B\"ulthoff concluded from visual matching experiments of simulated translucent objects that the ``physics of translucency is simply too complex for the visual system to run the generative equations~\emph{in reverse} and estimate intrinsic physical parameters via inverse optics ... instead the visual system relies on simple image heuristics [cues] that correlate with translucency"~\cite{FlemingBuelthoff2005}. They have also investigated the impact of color on perceived translucency and came to the conclusion that ``the saturation component is neither necessary nor sufficient to yield an impression of translucency". This is a good reason to keep \A independent of chroma and hue. We note that later investigations~\cite{Xiao2012_abstract} suggested, qualitatively, that chromatic components may have an impact on translucency perception. Given the lack of quantitative results concerning this, we leave consideration of this for future work.

Gkioulekas \etal evaluated the impact of the phase function on the appearance of translucent materials~\shortcite{GkioulekasXiaoZhaoAdelsonZicklerBala2013}. The phase function describes the scattering-induced distribution of light within a material and in addition to the absorption and scattering coefficient it is one parameter of the Radiative Transfer Equation~\cite{Chandrasekhar1960} that describes light transport within materials. By linearly combining the Henyey-Greenstein (HG) and von Mises-Fisher (vMF) lobes, they created a family of phase functions that significantly extends the often used HG model. They showed by conducting extensive paired-comparison experiments on rendered objects and multidimensional scaling that the physical parameter space can be embedded into a two-dimensional meaningful appearance space. The two-dimensional embedding is robust against variations in the object's shape, scattering coefficient and lighting variations and indicates that translucency perception is low dimensional even though the family of investigated phase functions is likely not complete as stated by the authors.

Motoyoshi~\shortcite{Motoyoshi2010} investigated visual translucency cues by two visual experiments on computer simulated objects. He suggests that a robust cue for translucency is the high-spatial-frequency luminance contrast of the non-specular component. If lateral subsurface light transport increases, this contrast is reduced and an object is perceived as more translucent. This perceptual cue has already been exploited in 3D printing by using gray-scale edge enhancement to offset the translucency of printing materials and make geometric details more visible~\cite{Cignoni_2008}, albeit in an ad hoc way. For highly translucent or transparent objects most of the incident light that is not reflected at the surface (Fresnel reflection) passes through the object without being back reflected by scattering. This causes shadow regions of the object to become brighter (compared to an opaque material) because of internal light contribution, and directly illuminated areas to become darker (compared to an opaque material), because of reduced back reflection within the object. The result is a reversed luminance contrast. 

Based on the observations above, we design our definition of \A to control two physical phenomena closely linked to perceptual cues: lateral and vertical subsurface light transport. Here, lateral and vertical are designed in terms of the observer's viewpoint with respect to the object. Lateral (subsurface) light transport refers to the extent to which light arriving on surfaces facing the observer is scattered laterally (with respect to the surface normal) before exiting the object in the direction of the observer. Vertical light transport refers to the extent to which light arriving on surfaces facing away from the observer is transported forward into the object before exiting in the direction of the observer.

For \A to reflect this robust translucency cue, it must be linked to both lateral and vertical subsurface light transport. Our approach to define \A is based on a set of virtual \emph{reference materials}, which are defined by wavelength-independent absorption and scattering coefficients. These materials show various magnitudes of lateral and vertical subsurface light transport causing a wide range of translucency cues. To measure \A for other materials, we propose a simple setup to measure lateral and vertical subsurface light transport of these materials and assign them an \A that corresponds to a reference material producing most similar measurements. The use of a library of virtual materials as a mechanism for search-space reduction is an established technique in computer graphics, and has recently been used for planar texture enhancement in 3D printing~\cite{ElekSumin2017}.

The approach of using reference materials for defining \A has \revision{three} big advantages compared to defining it only by a measurement procedure. \revision{First, measurement devices can be calibrated to these reference materials (see Section \ref{sec:DefiningA:Linking}) making the definition of \A not only device-independent with respect to the output (3D printer) but also to the input (measurement device).} \revision{Second}, a psychometric function (see Section \ref{sec:PsychometricFunction}), which endows \A with perceptual uniformity, can be determined using the reference materials by a psychophysical experiment (see Section \ref{sec:PsychophysicalExperiment}). \revision{Third}, since there is a direct transformation from the attenuation coefficient of the reference materials to $\A$, print size effects can be considered by applying a simple transformation to $\A$, which in turn has the effect of rescaling the attenuation coefficient and thereby adjusting \A to the new print size (see Section \ref{sec:Resizing}). 

\section{Defining \A}
\label{sec:DefiningA}

Mapping complex subsurface light transport material properties, described by a BSSRDF, to a single perceptually-meaningful value \A introduces significant loss of information. Our goal is to ensure that a large fraction of the available perceptual translucency cues of a given material can be reconstructed from $\A$. 

Our approach is inspired by that of defining standard RGB color spaces~\cite{SuesstrunkBuckleySwen1999}, such as sRGB, Adobe RGB, etc., and to represent object reflectances within these spaces. To define standard RGB color spaces, display primaries are selected to span a distinct color gamut. In general, spectral stimuli resulting from object reflectances for typical viewing conditions cannot be created by display primaries. The assignment of object reflectances to RGB values is done by metameric matches and gamut mapping: the reflectance information relevant for our color vision is extracted and mapped to the set of colors spanned by the primaries. Note that many reflectances, called \emph{metamers}, are assigned to the same RGB value under specified viewing conditions.

To define \A we specify a set of reference materials parametrized by physical quantities. These materials have the role of the display primaries to define standard RGB color spaces, but instead of creating a desired color gamut by mixing the primaries, the parametrized reference materials give access to various magnitudes of lateral and vertical light transport for spanning a wide range of perceptual translucency cues. \revision{\A is analytically linked to the intrinsic optical properties of the reference materials considering a psychometric function optimized for perceptual uniformity.}.

\subsection{Reference materials}

To parametrize the reference materials, four intrinsic parameters are potential candidates since they all have an impact on light transport, modeled by the steady-state \emph{Radiative Transfer Equation} (RTE)~\cite{Chandrasekhar1960} with shape-dependent boundary conditions: the absorption and scattering coefficients, the phase function and the refractive index. Table \ref{Table::ReferenceMaterials} summarizes our choices for these quantities.

Reflection at the material-air-interface is determined by the refractive index (Fresnel reflection).
The larger the material's refractive index, the smaller is the critical angle above which the light is totally back reflected. Lateral light transport depends on the refractive index and the surface geometry, but there is no direct impact of the refractive index on vertical light transport for homogenous materials. Therefore, to minimize the impact of the refractive index on the light transport properties of the reference materials we set it to 1.3, which is similar to that of water and has the advantage that the critical angle of total reflection is large, i.e. the impact of total reflection at the material-air interface is small. Note that while 3D printing materials used in multimaterial printing typically have larger \revision{refractive} indices ($ \approx [1.4, 1.6]$), we can account for this discrepancy by Saunderson correction~\shortcite{Saunderson1942}, which is used in the coating industry before applying the Kubelka-Munk model~\shortcite{KubelkaMunk1931} to predict the coating's color. \revision{In this paper and for creating the prints \cite{BruntonArikanTanksaleUrban2018}, Saunderson correction was not used.}

Absorption has no impact on the direction of light transport and a good portion of absorption information is already included in the RGB dimensions. Nevertheless, it strongly influences the magnitude of light transport and cannot be neglected for defining $\A$. 
Following Fleming and B\"ulthoff~\shortcite{FlemingBuelthoff2005}, who found that saturation is not necessary to yield the impression of translucency, we consider absorption for our parametrization to be wavelength independent.

The scattering coefficient and the phase function have the largest impact on the direction of light transport. The average distance a photon travels before a scattering or absorption event occurs is called the \emph{mean free path} and is given by $1/(\absorp +\scatt) = 1/\atten$, where $\absorp$ is the absorption, $\scatt$ is the scattering and $\atten = \absorp +\scatt$ is the attenuation coefficient. Since \A has to approximate materials with all optical thicknesses from fully transparent ($\absorp = \scatt = \atten = 0$) to fully opaque ($\atten = \infty$), the scattering coefficient must also be considered for parametrizing our reference materials.

The phase function describes the angular distribution of photon directions after a scattering event. Gkioulekas~\etal showed that the phase function has an significant impact on the appearance of translucent materials~\shortcite{GkioulekasXiaoZhaoAdelsonZicklerBala2013} and presented a two-dimensional appearance space in which many (but likely not all) phase functions can be embedded. An optimal choice for our purpose would be the \emph{maximum likelihood phase function} minimizing the expected average perceptual error using the embedding and distance metric given by Gkioulekas~\etal Unfortunately, in contrast to the distribution of reflectances~\cite{AttewellBaddeley2007}, the distribution of phase functions within the visual environment was not investigated so far. Without this knowledge we have to assume a phase function distribution, which would make any maximum likelihood choice biased by this guess. 

For this reason, we use a heuristic choice of an isotropic phase function to avoid large errors since materials and material compositions may be dominated by forward scattering, such as skin~\cite{NaitoYamadaOgawaTakata2010}, or backward scattering, such as marble. This choice has another advantage related to linking the reference materials to a perceptual scale: Xiao~\etal~\shortcite{XiaoWalterGkioulekasZicklerAdelsonBala2014} reported that the human visual system (HVS) is able to estimate translucency in a consistent way across different shapes and lighting conditions only for simple materials, particularly those with an isotropic phase function. Reference materials possessing \emph{translucency constancy} make the derived perceptually uniform scale less dependent on lighting conditions and shape and thus more general. 

We do not claim that this choice of reference material parameters is optimal for spanning the maximum range of translucency cues. However, we show in Section \ref{Subseq::Measurements} that a wide range of lateral and vertical light transport magnitudes is covered.

\begin{table}
	\centering
		\caption{Reference material parameters. Absorption and scattering coefficients are wavelength-independent.}
		\begin{tabular}[tbh] {c c c c}
		Refractive index & Phase function & $\absorp$ & $\scatt$ \\
		\hline
		1.3 & isotropic & $[0,\infty)\cm^{-1}$ & $[0,\infty)\cm^{-1}$
		\end{tabular}
		\label{Table::ReferenceMaterials}
\end{table}

\subsection{Linking \A to reference materials}

\revision{Given a set of reference materials \refmats, w}e link \A to \revision{their} absorption and scattering coefficients $(\absorp,\scatt)$ as follows:
\begin{equation}
\A_{\refmats}(\absorp,\scatt) = \psychA(\Ahat_{\refmats}(\psychatten(\absorp,\scatt))) = \psychA(1-\vec{e}^{-c\psychatten(\absorp,\scatt)})
\label{Eq::A2ReferenceMaterials}
\end{equation}
where $\psychatten(\absorp,\scatt) \in [0,\infty)$ is a modified attenuation coefficient considering that absorption and scattering may have a different impact on translucency perception, which is determined in Section \ref{sec:PsychometricFunction}. $\Ahat_{\refmats}$ is an intermediate value, which depends only on the modified attenuation coefficient. It is zero for the totally transparent reference material with $\psychatten(0,0) = 0 \cm^{-1}$, whereas for $\psychatten \rightarrow \infty \Rightarrow \Ahat_{\refmats} \rightarrow 1$. 
The constant $c = 0.0153 \cm$ is used to make $\Ahat_{\refmats}$ dimensionless and to scale $\Ahat_{\refmats}$ to 0.99 for $(\absorp,\scatt) = (0,300) \cm^{-1}$ for the case $\psychatten(\absorp,\scatt) = \atten  = \absorp + \scatt$. Such scaling ensures that a large portion of the $\A$-range is exploited by real materials. Inspired by Lambert's law, we use a negative exponential function to create a linear relationship between $\Ahat_{\refmats}$ and the magnitude of light transport within the reference materials. Lambert's law breaks for highly scattering media, and the relationship between $\Ahat_{\refmats}$ and the magnitude of light transport will therefore likely deviate from linearity for such reference materials. Finally we use a psychometric function \psychA to convert $\Ahat_{\refmats}$ into a perceptually more uniform representation that minimizes the disagreement between perceived translucency distances and Euclidean distances in $\A_{\refmats}$. The \emph{translucency constancy} for the reference materials generalizes this psychometric function also to other lighting conditions and shapes~\cite{XiaoWalterGkioulekasZicklerAdelsonBala2014}. In Section \ref{sec:PsychometricFunction} we describe a simple model for $\psychA$ based on Stevens' power law~\cite{Stevens1960}.

\revision{Note that $\A_{\refmats}$ is dependent on and determined by \refmats, and is not defined in its absence. In particular, $\A=\A_{\refmats}$ for any material $M$, including $M\notin\refmats$, as we see in Section \ref{Subseq::Measurements}. Since in this paper we use a fixed set of reference materials, we drop the subscript to ease the notation. However, considering different sets of reference materials for different objects makes for an interesting avenue for future work.}

\begin{figure}[tbh]
\centering
\includegraphics[width=0.45\textwidth]{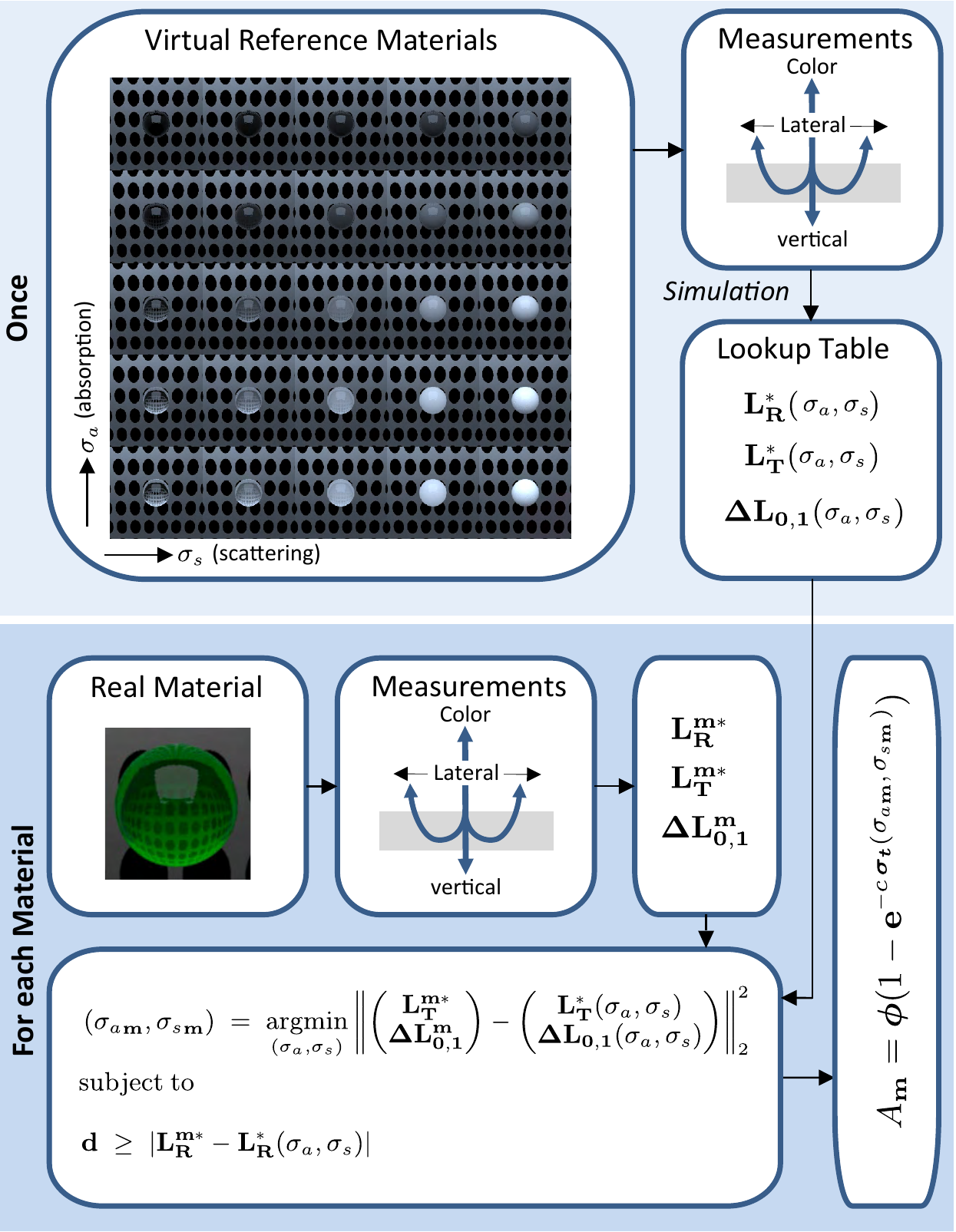} 
  \caption{Concept of obtaining a real material's \A value via reference materials and light transport measurements}
	\label{Figure::DefiningA}
\end{figure}

\begin{figure*}[tbh]
\includegraphics[width=\textwidth]{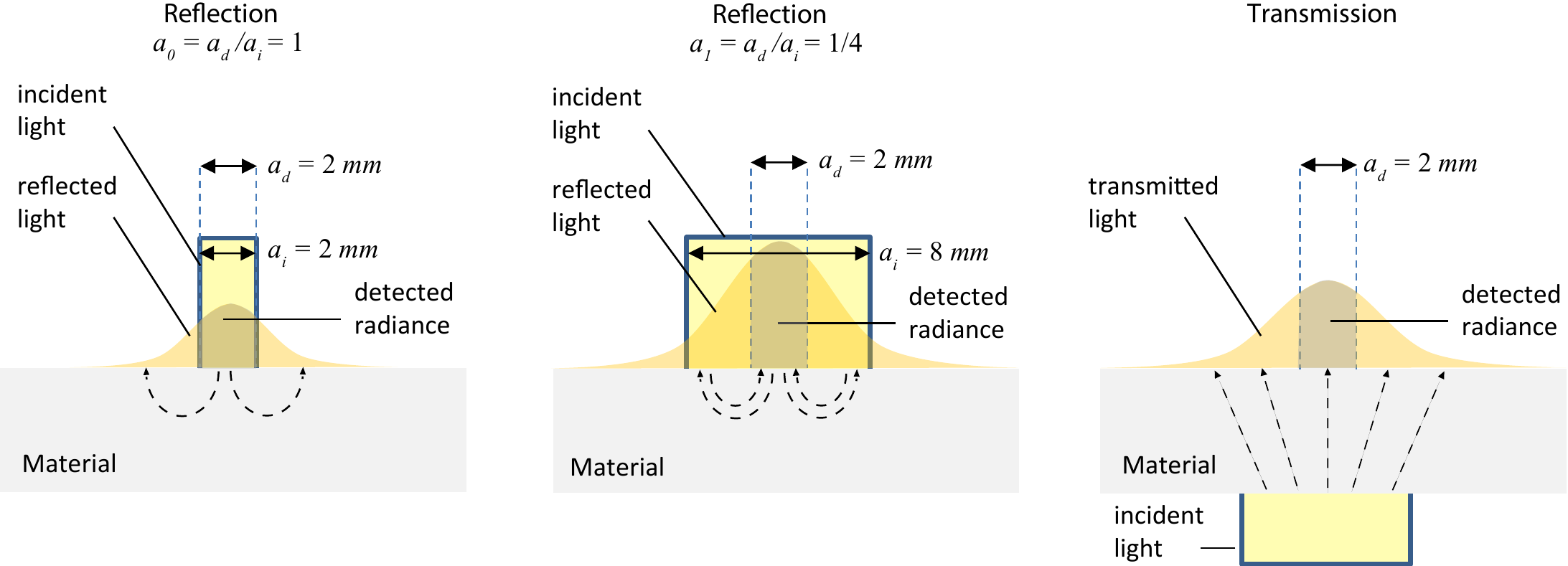} 
  \caption{Reflectance measurements for $a_0 = 1$ (left) and $a_1 = 1/4$ (middle) condition. Due to subsurface light transport only away from the detection area the edge-loss for $a_0$ is larger than for $a_1$. Vertical light transport is measured by a transmission setup (right), lateral by the edge-loss difference.}
	\label{Figure::EdgeLoss}
\end{figure*}

\section{Measuring \A}
\label{Subseq::Measurements}

\revision{We measure \A with an indirect approach facilitated by the reference materials. Given an unknown material, we begin by measuring its lateral and vertical light transport, subject to color measurement conditions, and then find the reference material with the best matching lateral and vertical light transport measurements (simulated once for all reference materials in a pre-process). From the resulting reference material, we have the necessary scattering and absorption coefficients to compute \A via \eqref{Eq::A2ReferenceMaterials}. Figure \ref{Figure::DefiningA} shows this process.}

\subsection{Measuring lateral and vertical light transport}
Various approaches were proposed to phenomenologically measure light transport by projecting a structured light pattern onto the material and recover the BSSRDF or the point spread function (lateral light transport) from the captured response signal~\cite{GoeseleLenschLangFuchsSeidel2004,PeersBergeMatusikRamamoorthiLawrenceRusinkiewiczDutre2006,YuleNielsen,UkishimaKanekoNakaguchiTsumuraHautaKasariParkkinenMiyake2009,HappelDoersamUrban2014}. Recent approaches use computational imaging techniques for inferring the intrinsic RTE parameters~\cite{GkioulekasLevinZickler2016}. Even though these measurement approaches are time consuming and work so far only in a laboratory environment, commercial devices might be available in future. Nevertheless, a simple, fast and ideally already commercially available measurement setup is desired for practical applications. Such setup needs to measure lateral and vertical light transport to extract only the BSSRDF information responsible for creating the perceptual translucency cues, i.e. impacting the high-spatial-frequency luminance contrast~\cite{FlemingBuelthoff2005,Motoyoshi2010}.

To specify lateral light transport we propose to use the spectrophotometric edge-loss difference $\mathbf{\Delta L_{0,1}} = |\mathbf{L_0^*}-\mathbf{L_1^*}|$, where $\mathbf{L_0^*}$ and $\mathbf{L_1^*}$ are lightness values obtained from reflectance measurements of the same sample employing two different measurement conditions~\cite{YoshidaKomedaOjimaIwata2011}. A measurement condition is characterized by the ratio $a = a_d/a_i$, where $a_d$ is the diameter of a circular detection area and $a_i$ is the diameter of a circular illuminated area used by the spectrophotometer. For $a \approx 1$, the edge loss is large because incident light is transported within the material away from the detection area and cannot contribute to the measurement. The edge loss becomes smaller with decreasing $a$ because the illuminated area is larger than the detection area and light is also transported towards the detection area reducing the edge loss. Figure \ref{Figure::EdgeLoss} illustrates edge-loss for two measurement conditions. The measurement differences between the two conditions with different edge loss magnitudes is a quantity characterizing lateral light transport. Note that spectrophotometric measurements are always relative to a reference measurement and factors such as the sensor's quantum efficiency or the light source's \emph{Spectral Power Distribution} (SPD) are canceled out. More details including a derivation of the measurement equations from BSSRDFs is given by Yoshida~\etal~\shortcite{YoshidaKomedaOjimaIwata2011}.

We measure vertical light transport using a transmission setup, where the material is placed between detector and illumination (see Figure \ref{Figure::EdgeLoss} right). From the recorded SPD of the light transmitted through the material, we compute the lightness $\mathbf{L_T^*}$ relative to the fully transparent material (air). 

To be consistent with the color information, we suggest to compute lightness from reflectance or transmittance spectra using the CIE 1931 observer and the CIE D50 illuminant as specified by the \emph{International Color Consortium} (ICC)~\cite{ICC43}.

Commercial reflectance spectrophotometers used in graphic arts employ a circular 45/0 measurement geometry. For our purpose this off-specular measurement geometry is favorable because the extracted BSSRDF information does not interfere with surface reflection which is independent of light transport and does not contribute to the translucency cues~\cite{Motoyoshi2010}.

A spectrophotometer with the desired capabilities is the Barbieri Spectro LFP. It can be used in transmission (d/0 geometry) and reflection (circular 45/0) mode and allows to measure materials up to a thickness of $2 \cm$. For the edge-loss measurements, we use $a_i = 2 \mm$ and $a_i = 8 \mm$ aperture for illumination with a constant detection aperture of $a_d = 2 \mm$, i.e. we can run it with the condition $a_0 = 2/2 = 1$ and $a_1 = 2/8 = 1/4$. We select the thickness of the material samples to be $0.4 \cm$. A black backing is used for reflectance measurements to ensure that light emitted on the other side of the material sample is not contributing to the reflectance measurments, which might bias the edge loss. Table \ref{Table::MeasurementConditions} summarizes the proposed measurement conditions employing the Barbieri Spectro LFP. The meaningfulness of the edge-loss measurements w.r.t. the HVS is demonstrated in Appendix \ref{sec:meaningful}.

\begin{table}
	\centering
		\caption{Measurement conditions}
		\begin{tabular}[tbh] {r c} 
		\textbf{Edge-Loss reflectance (lateral)} & \\ \hline
			 Geometry & 45/0 \\
			 Backing  & black\\ \hline			 
		   Condition $a_0 = 1$:  & \\ 		
			 Detection diameter $a_d$ & $2 \mm$  \\
			 Illumination diameter $a_i$ & $2 \mm$  \\ \hline
 		   Condition $a_1 = 1/4$: & \\ 		
			 Detection diameter $a_d$ & $2 \mm$  \\
			 Illumination diameter $a_i$ & $8 \mm$  \\ \hline			
		 \textbf{Transmittance (vertical)} & \\ 
				 Geometry & d/0 \\ \hline
		 \textbf{Reflectance (RGB)} & \\
				Standard & ISO13655 \cite{ISO13655}\\ 
				Backing & white \\ \hline				
				Patch Thickness & $4 \mm$ \\ \hline
		\end{tabular}
		\label{Table::MeasurementConditions}
\end{table}

\subsection{Specifying color measurements conditions}
\label{Seq::SpecfyingColorMeasurementConditions}
Multiple reference materials may result in the same \A value because definition \eqref{Eq::A2ReferenceMaterials} combines absorption $\absorp$ and scattering $\scatt$ coefficient in the modified attenuation coefficient \psychatten. Therefore, information included in the RGB data is necessary to retrieve the reference material from $\A$. Since the saturation component is not necessary for the perception of translucency~\cite{FlemingBuelthoff2005}, we use lightness $\mathbf{L_R^*}$ that can be extracted from the RGB signal by converting the standard RGB values to CIEXYZ and then to CIELAB. To avoid inconsistencies between RGB and \A when retrieving the reference material, we must ensure that the lightness of the reference material is similar to the lightness of the RGB values measured under the same conditions.

These conditions are given in ISO13655~\cite{ISO13655} and supported by all spectrophotometers used in graphic arts. ISO13655 makes provision for either black or white backing. In order to support also optically thin materials, we must resolve this ambiguity for not obtaining extremely different reflectances depending on the choice of the backing. 

We suggest to use white backing to be consistent with measurement conditions recommended for color-only 3D printing, where the core of the print is filled with white material to maximize reflectance if highly translucent materials are used~\cite{BruntonArikanUrban2015}. For this reason, white backing agrees more than black backing with the material arrangement conditions contributing to the final print's color~\cite{ArikanBruntonTanksaleUrban2015}.


\subsection{Linking measurements to reference materials}
\label{sec:DefiningA:Linking}

\revision{Our measurement setup does not directly give us the necessary quantities to compute $\A$. However, for each reference material we have $\A$, and we can conduct virtual measurements according the above setup via simulation.}

We simulated Barbieri LFP transmittance lightness and reflectance edge-loss difference measurements for reference materials with all combinations of the following scattering and absorption coefficients: $\{$0, 0.05, $\dots$, 0.8, 0.9, $\dots$, 1.4, 1.6, $\dots$, 2, 4, $\dots$, 10, 20, 50, 75, 100, 200, 300, 600, 850, 1000, 1250, 2500$\} \cm^{-1}$. For this, we used CAD data of the spectrophotometer's optical path and solved the full steady-state RTE using Mitsuba's Monte-Carlo path tracer~\cite{Mitsuba}. Employing Mitsuba, we simulated also the color of the samples for measurement conditions specified in section \ref{Seq::SpecfyingColorMeasurementConditions} and computed the reflectance lightness. 

\begin{figure*}[tbh]
\includegraphics[width = \textwidth]{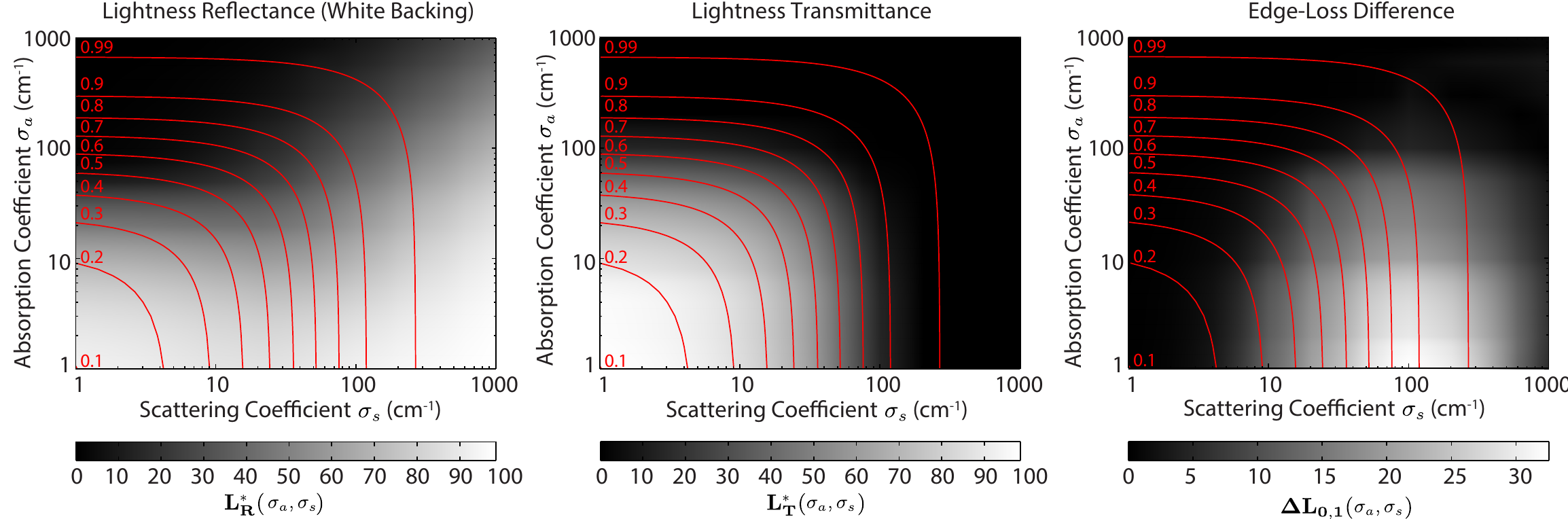} 
  \caption{Light transport measurements simulated for the reference materials as a log-log plot of absorption and scattering coefficients. The red curves indicate constant \A employing the simple models for \psychatten and \psychA as suggested in Section \ref{Sec:Simplemodels}.}
	\label{Figure::SimulationsReferenceMaterialsAndA}
\end{figure*}

By bilinearly interpolating intermediate values of simulated measurements corresponding to absorption and scattering coefficients, we define functions that map every $(\absorp,\scatt) \in [0,2500]^2 $ to reflectance lightness $\mathbf{L_R^*}(\absorp,\scatt)$, transmittance lightness $\mathbf{L_T^*}(\absorp,\scatt)$ and reflectance edge-loss difference $\mathbf{\Delta L_{0,1}}(\absorp,\scatt)$. Light transport quantities of the reference materials are shown in Figure \ref{Figure::SimulationsReferenceMaterialsAndA}.

For obtaining $\A_\mathbf{m}$ of an arbitrary real material, we measure color and light transport quantities $(\mathbf{L_R^{m*}}$, $\mathbf{L_T^{m*}}$, $\mathbf{\Delta L_{0,1}^{m}})$ on patches with the same thickness as the patches used for the simulation and solve the following optimization problem
\begin{eqnarray}
(\absorp_\mathbf{m},\scatt_\mathbf{m}) & = & \argmin_{(\absorp,\scatt)} 
\left\|
\begin{pmatrix}
	\mathbf{L_T^{m*}} \\
	\mathbf{\Delta L_{0,1}^{m}}	
\end{pmatrix}-
\begin{pmatrix}
	\mathbf{L_T^*}(\absorp,\scatt) \\
	\mathbf{\Delta L_{0,1}}(\absorp,\scatt)	
\end{pmatrix}
 \right\|_2^2 
\label{Eq::Distance} \\
\mathrm{subject \ to \ } \vec{d} & \geq & |\mathbf{L_R^{m*}} - \mathbf{L_R^*}(\absorp,\scatt)|, \nonumber 
\end{eqnarray}
where $\vec{d} > 0$ is an acceptance threshold for inconsistency between the lightness of the RGB color and the lightness of the reference material. We used $\vec{d} = 2$ in this paper. To solve optimization problem \ref{Eq::Distance}, we used exhaustive search discretizing the 2D absorption-scattering space of reference material in 0.1 $\cm^{-1}$ units. We then obtain the material's $\A_\mathbf{m}$ value from the \revision{reference material's} $(\absorp_\mathbf{m},\scatt_\mathbf{m})$-\revision{coefficients} using Eq. \eqref{Eq::A2ReferenceMaterials}.

It is worth mentioning that lightness and lightness difference units belong to a nearly perceptually-uniform lightness scale, i.e. the contribution of the two addends of the objective function \eqref{Eq::Distance} are perceptually balanced. Figure \ref{Figure::DefiningA} illustrates the \revision{process of using virtual measurements of references materials to determine $\A$ for a real material. Again, note that the definition of \A for \emph{any} material is dependent on the set of reference materials \refmats. Therefore, although the scattering and absorption coefficients obtained by \eqref{Eq::Distance} differ in general from those of the real material, the value obtained for $\A_\mathbf{m}$ is that of the real material, given \refmats.}  

Note that various materials with different BSSRDFs might have the same \A value. Information loss occurs if lateral and vertical light transport of a material are not well represented by any reference material. In this case, the optimal reference material according to Eq. \eqref{Eq::Distance} under- or overestimates lateral or vertical light transport and produces deviating translucency cues. Such materials usually posses non-isotropic phase functions or wavelength-dependent absorption or scattering. Examples are shown in Figures \ref{Figure::Dragon}-\ref{Figure::Temple}.

Given an RGBA value, we can retrieve the reference material's $(\absorp,\scatt)$ values, by computing 
\begin{eqnarray}
(\absorp,\scatt) & = & \argmin_{(\absorp',\scatt')} |\mathbf{L_{\mathrm{\tiny RGBA}}} - \mathbf{L_R^*}(\absorp',\scatt')|
\label{Eq::FindingReferenceMaterial} \\
\mathrm{subject \ to \ } \A(\absorp,\scatt) & = & {\A}_{\mathrm{\tiny RGBA}}, \nonumber
\end{eqnarray}
where $\mathbf{L_{\mathrm{\tiny RGBA}}}$ is the reflectance lightness obtained from RGB. 
The solution of problem \eqref{Eq::FindingReferenceMaterial} has the smallest lightness difference to the RGB lightness and lies on a curve corresponding to constant $\A_{\mathrm{\tiny RGBA}}$. Figure \ref{Figure::SimulationsReferenceMaterialsAndA} (left) illustrates some curves of constant $\A$. We precompute a lookup table to solve \eqref{Eq::FindingReferenceMaterial} in $O(1)$ time.

\begin{figure}[tbh]
\centering
\includegraphics[width=0.45\textwidth]{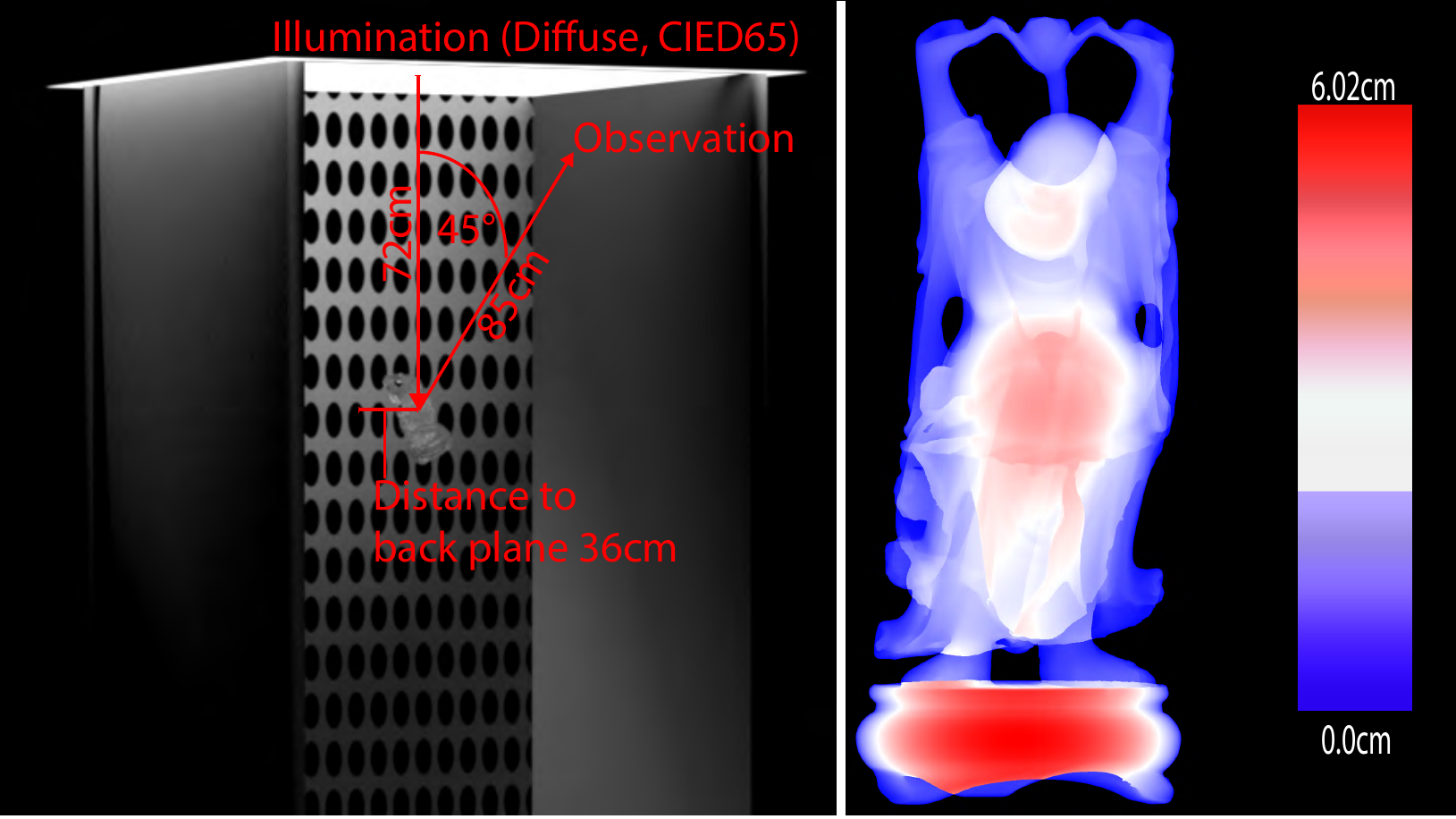} 
  \caption{Rendering conditions and sample thickness for the experiment.}
	\label{Figure::SampleAndGeometry}
\end{figure}

\section{Psychophysical Experiment}
\label{sec:PsychophysicalExperiment}

Psychophysical experiments are required to obtain perceptual translucency differences used to determine \psychatten and \psychA for minimizing the disagreement between perceptual differences and Euclidean distances in $\A$.

\subsection{Stimuli}

\subsubsection{Perceptual uniformity of color spaces and the choice of the psychophysical method}
Psychophysical experiments show that a perceptually uniform color space does not exist~\cite{MacAdam1963}: Let $x$ and $y$ be two colors in a color space. If we chose a third color $z$ lying on the geodesic curve with respect to threshold differences and cutting it into two equal parts so that the distances $\overline{xz}$ and $\overline{zy}$ are perceived equally, then the distance $\overline{xy}$ is perceived smaller than $\overline{xz} + \overline{zy}$. The biggest implication of this \emph{diminishing returns on color difference perception} effect is that a color space can only be close to perceptual uniformity for a distinct magnitude range of color differences but not across color difference ranges. Since in most applications suprathreshold small color differences, slightly above the just noticeable difference (JND), are important, color spaces such as CIELAB, are designed to be nearly perceptually uniform only for small color differences. Also color difference formulas, such as CIE94 or CIEDE2000, to improve the uniformity of CIELAB are based on experiments for small color differences. 
\revision{Note that the Gaussian curvature calculated from line elements fitted to various visual data of small color differences is nonzero \cite{WyszeckiStiles2000}. This indicates that an error-less Euclidean embedding in three dimensions is impossible and that a perfectly perceptually-uniform color space does not exist.}

We are not aware of visual experiments investigating the existence of \emph{diminishing returns on translucency difference perception}. If such an effect exists, a Euclidean embedding based on large translucency differences employing, for instance, multidimensional scaling would deviate from an embedding based on small translucency differences. Similar to color differences, most applications such as fine-tuning translucency in 3D printing, would benefit from an Euclidean embedding based on small translucency differences. Therefore, we use the \emph{method of constant stimuli}~\cite{Engeldrum2000} employing an anchor pair with a translucency difference slightly above threshold for obtaining small visual translucency differences. Stimuli are shown on a color calibrated display and were computed using the Monte-Carlo path tracer Mitsuba~\cite{Mitsuba} solving the steady-state RTE for each wavelength of the visible wavelength range sampled in 10$\nm$ steps. 

\subsubsection{Virtual Viewing Booth}
A virtual viewing booth with a dimension of depth x width x height  = $72\cm \times 72\cm \times 144\cm$ was designed employing a perfectly-diffuse area light emitter covering the whole ceiling of the booth. The spectral power distribution of the light is set to CIE D65 (daylight). The remaining parts of the booth have a gray value of CIE L$^* = 79$ and the back plane is covered with black circles, giving the observer additional hints for judging translucency, particularly in the optically thin range.

\subsubsection{Sample and Viewing Geometry}
Perceived translucency strongly depends on the shape -- particularly the thickness -- of an object. We used the Stanford Happy Buddha model\footnote{https://graphics.stanford.edu/data/3Dscanrep/} with dimensions of depth x width x height $6.5\cm \times 6.5\cm \times 15.8\cm$. This model covers various surface orientations and thickness levels as shown in Figure \ref{Figure::SampleAndGeometry}. It also contains many thin regions that are particularly important for the human visual system to judge the degree of translucency~\cite{FlemingBuelthoff2005,NagaiOnoTaniKoidaKitazakiNakauchi2013,XiaoWalterGkioulekasZicklerAdelsonBala2014}. The model is placed into the booth so that its center has a distance of $36\cm$ from the back plane and $72\cm$ from the ceiling. The model's length axis is rotated so that the model's front side is illuminated from approx. 45$^\circ$. Renderings are computed from a viewpoint perpendicular to the model's front side.

\subsubsection{Sample selection}
\begin{figure}
\centering
\includegraphics[width=0.48\textwidth]{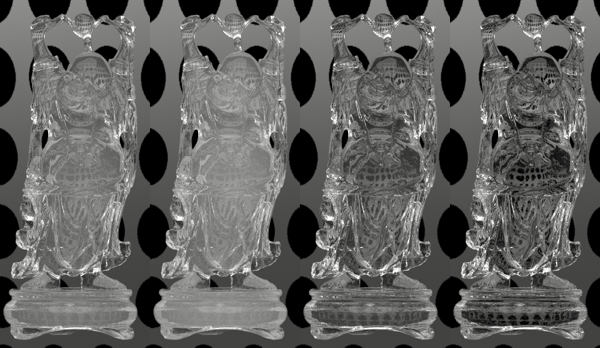} 
  \caption{Example images shown to subjects: Center/test pair (left) and anchor pair (right)}
	\label{Fig:BuddhaExptExample}
\end{figure}

For the method of constant stimuli two pairs of samples are shown to subjects simultaneously: An anchor pair and a test pair. The anchor pair consists of two samples with a suprathreshold translucency difference (see Figure \ref{Fig:BuddhaExptExample}) and the scattering coefficients $0 \cm^{-1}$ and $1.5 \cm^{-1}$. The test pair consists of a center sample and a test sample whose translucency difference is compared in a trial to the anchor pair (see section \ref{Subsubsection::Procedure}). 

A first set of test pairs consists of samples with zero absorption: In a preliminary experiment, we selected 6 center samples at scattering coefficients $\{0, 4.5, 12, 40, 75, 300\} \cm^{-1}$. The perceived translucency difference between the center samples was selected to be clearly larger than the one of the anchor pair. For each center sample, we selected 7 test samples of increasing and 7 test samples of decreasing scattering coefficients. For the center sample with scattering coefficient $0 \cm^{-1}$ only test samples of increasing and for center sample with scattering coefficient $300 \cm^{-1}$ only test samples with decreasing scattering coefficients are used. We ensured that the maximum perceived difference between center sample and test sample in each direction is clearly larger than the one of the anchor pair and the minimum difference is clearly smaller. Each subject was shown a total of 6 (center samples) x 2 (directions) x 7 (test samples) - 2 (directions) x 7 (test samples) = 70 center/test pairs next to the anchor pair. 

A second set of test pairs consists of samples with nonzero absorption. To make the results more robust to chromatic variations, we selected reference materials, but replaced the wavelength-independent absorption spectra by red, green and blue spectra, resulting in samples colored in the primaries of an RGB space. The anchor pair was always compared to a test pair possessing the same absorption spectrum and varying only in scattering, where scattering coefficients were selected in a similar way as for the first set of pairs. Note that the reflectance lightness values of these materials when measured according to Section \ref{Subseq::Measurements} almost uniformly cover the range from $\mathbf{L_R^{m*}} = 1$ to $\mathbf{L_R^{m*}} = 80$. For each of the three absorption spectra, a subject was shown 5 (center samples) x 2 (directions) x 6 (test samples) - 6 (directions) x 2 (test samples) = 58 center/test sample pairs next to the anchor pair, resulting in a total of 174 comparisons. See Appendix \ref{sec:color_test} for more details.

The position (left/right) of the anchor and test pairs was randomized during the experiment to minimize any sequential effects.

\subsubsection{Viewing Conditions}
\label{sec:PsychophysicalExperiment:viewcond}
The experiment was conducted in a dark room with no other source of illumination other than the display. Two color calibrated displays, the EIZO ColorEdge CG301W and CG276, were used in the experiments. The displays were calibrated to a white point of CIE D65. At the beginning of the experiment, the luminance of the rendering of a white opaque patch (perfectly reflecting diffuser) placed at the same position as the model was measured and a scaling factor was computed to set the patch's luminance to $190 \unit{cd/m}^2$. The luminance was directly measured on the displays with a Konica Minolta CS-1000A and a Topcon SR-3AR spectroradiometer. All renderings were than normalized using this scaling factor. 
For the method of constant stimuli, four $5\cm$ wide renderings (anchor pair, center/test pair) are shown next to each other centered on the display. The remaining display area is set to the same gray color as the back panel of the virtual viewing booth. In the experiment, the subjects had a viewing distance of approx. $60\cm$ from the display so that each image occupied approx. 4.8$^\circ$ and each model 4.24$^\circ$ of the visual field. For the original $6.5\cm$ wide model, 4.24$^\circ$ of the visual field corresponds to a viewing distance of approx. $88\cm$. 

\subsection{Experiments}
\subsubsection{Subjects}
A total of 40 subjects, 6 females and 34 males all with normal visual acuity according to the Snellen test participated in the experiment. All subjects passed the Ishihara color deficiency test. Their average age was 25.67 years with a standard deviation of 3.81 years.

\subsubsection{Procedure}
\label{Subsubsection::Procedure}
The anchor pair was shown with a center/test pair in random order on the display. The subjects were asked to select the pair that has the larger apparent translucency difference. No explicit definition of translucency was given to subjects but they were asked to not judge the difference of a pair based on color or overall lightness. 

\begin{figure}[tbh]
\includegraphics[width = 0.45 \textwidth]{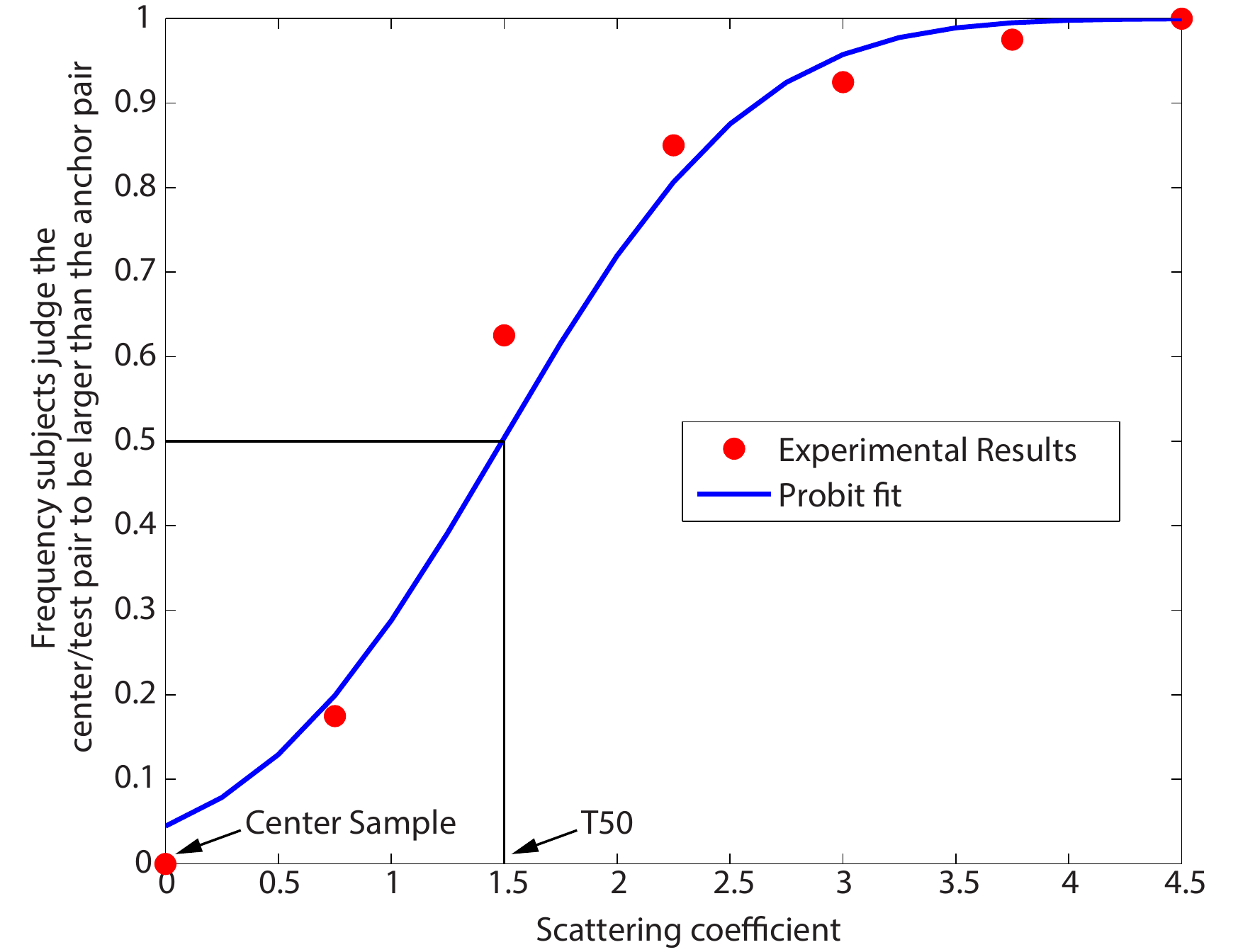} 
  \caption{Probit fit}
	\label{Figure::ProbitFit}
\end{figure}

\section{Psychometric Function and Modified Attenuation}
\label{sec:PsychometricFunction}

\subsection{Estimating visual translucency differences}
For precise estimates of population tolerances, intra-observer uncertainty must be minimized. For this, we applied a low-pass filter algorithm on each observer's binary data to reduce non-monotonic responses for each direction as suggested by Berns \etal \shortcite{BernsAlmanReniffSnyderBalononRosen1991}. We used Probit analysis on the resulting data to estimate the so called T50 distances from the selected sample centers. The T50 distance is the translucency difference from the sample center that is judged by 50\% of the population to be smaller than the perceived translucency difference of the anchor pair and to be larger by the remaining 50\% of the population. 

Only T50 values for which the probit fit passed the $\chi^2$ goodness-of-fit test ($\alpha = 0.05$) were considered for the following evaluation. For the zero-absorption samples 8 out of 10, for the red and green 6 out of 8 and for the blue 4 out of 8 passed this test. A total of 24 visual translucency differences were considered. Rejections were characterized by a test sample choice spanning a too small visual translucency difference interval so that many observers found all test pairs in this direction to have a smaller translucency difference than the anchor pair. 

For the second set of test pairs with red, green and blue absorption spectra, we computed for each material corresponding to a center sample and to the T50 distance the reference material as described in Section \ref{sec:DefiningA:Linking} using simulated measurements.

Figure \ref{Figure::ProbitFit} illustrates the Probit analysis for the transparent center sample. All data is provided in the supplementary material.

\subsection{Simple models for the modified attenuation coefficient and the psychometric function}
\label{Sec:Simplemodels}
We aim to find simple models for \psychatten and \psychA to obtain a perceptual uniformity for translucency comparable with the color uniformity of CIELAB. For this, we use a simple linear model for $\psychatten_p(\absorp,\scatt) = p\absorp + \scatt$ and Stevens' power law for $\psychA_q(\Ahat) = \Ahat^q$, yielding a two parameter model $\A_{r}(\absorp,\scatt), \ r = (p,q)$, if inserted into Eq. \eqref{Eq::A2ReferenceMaterials}.

Since all visual translucency differences obtained in the visual experiment match with the anchor pair's translucency difference, we can fit the model parameters $r = (p,q)$ by minimizing the disagreement of distances:

\begin{eqnarray}
r' & = & \argmin_{r} \sum_{t \in \vec{V}} \left(\Delta\A_{r}(t) - \Delta\A_{r}(t_a)\right)^2 \label{eq:Optimization}
\end{eqnarray}
where $\vec{V}$ is the set of the 24 different absorption and scattering pairs obtained in the visual experiment corresponding to a perceived translucency differences similar to the one of the anchor pair, $\Delta\A_{r}(t) = |\A_{r}(\absorp_{t1},\scatt_{t1}) - \A_{r}(\absorp_{t2},\scatt_{t2})|$ is the distance in $\A_{r}$ for the absorption and scattering pair $t = ((\absorp_{t1},\scatt_{t1}),(\absorp_{t2},\scatt_{t2})) \in \vec{V}$, and $t_a \in \vec{V}$ corresponds to the anchor pair.

To evaluate the performance of the fit, we use the \emph{STandardized REsidual Sum of Squares} (STRESS) index (see Appendix \ref{Appendix:STRESS}). In our case the predicted difference is always $\Delta\A_{r}(t_a)$ and the visual translucency differences are $\Delta\A_{r}(t), \in \vec{V}$. The STRESS index allows a significance comparison using the F-test. 

\begin{figure}
\centering
\includegraphics[width=0.45\textwidth]{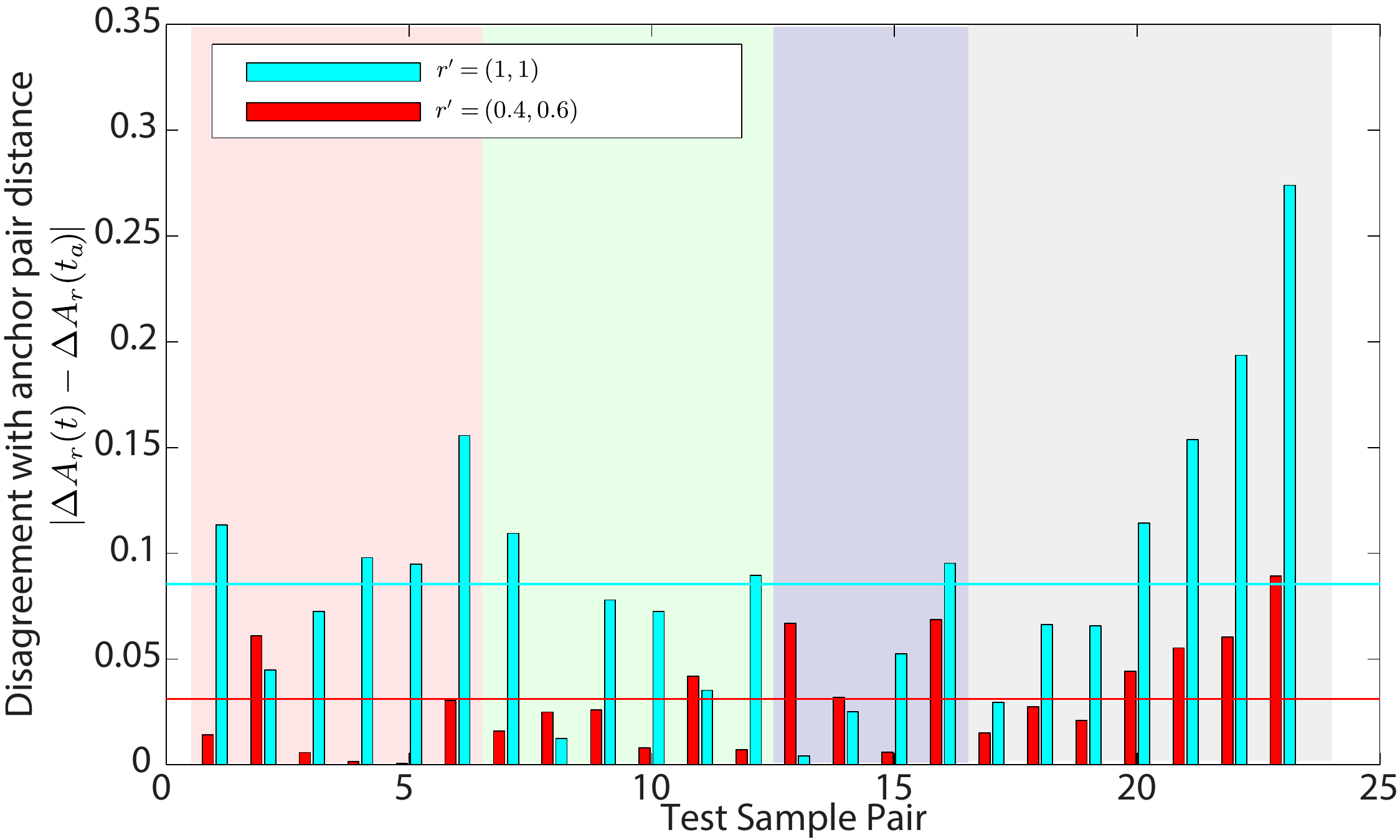} 
  \caption{Disagreement between $\A_r$ distances and anchor pair distance on the visual data: Using simple models for psychometric function and modified attenuation (red bars, average = horizontal red line) and not considering these models (cyan bars, average = horizontal cyan line). The background color indicates the test sample set (red, green, blue and no absorption).}
	\label{Figure::ADisagreement}
\end{figure}

The result of optimization \eqref{eq:Optimization} is $r' = (p',q') = (0.4,0.6)$ with a STRESS value of 32,7. The numbers are already rounded to one decimal place without significantly changing the uniformity according to the F-test. Constant \A values employing this model are shown in Figure \ref{Figure::SimulationsReferenceMaterialsAndA}.
For the not modified attenuation coefficient and without applying the psychometric function, i.e. $r' = (p',q') = (1,1)$, the STRESS value is 49.7. According to the F-test the uniformity for $r' = (0.4,0.6)$ is significantly better than for $r' = (1,1)$. Figure \ref{Figure::ADisagreement} compares the performance of both models \revision{and Appendix \ref{sec:PerceptuallyUniformityAAhat} shows renderings of Lucy corresponding to a uniform sampling of both models for reference materials covering the visually-relevant absorption and scattering range}. It is noteworthy that by non-linearly scaling the absorption parameter $\psychatten_(p_1,p_2)(\absorp,\scatt) = p_1\absorp^{p_2} + \scatt$ the STRESS value would further drop to 26.6. However, this would not significantly improve uniformity according to the F-test, it would require another parameter and it would break the modified attenuation coefficient's unit. For this reason, we suggest to use the linear model for \psychatten.

We also analyzed whether our model is overfitting the data using an exhaustive leave-one-out cross-validation test. The disagreements for the left-out visual translucency differences are: (mean, std, max) = (0.0352, 0.0255, 0.0857) and almost similar to the disagreements for all visual translucency differences (mean, std, max) = (0.0302, 0.0226, 0.0764). Also the model parameters are quite robust: (min$_p$, mean$_p$, max$_p$) = (0.31, 0,41, 0.47) and (min$_q$, mean$_q$, max$_q$) = (0.57, 0.58, 0.59), which suggest that the model does not overfit the data.

To put the STRESS index of 32.7 into perspective: The nearly-perceptually uniform CIELAB color space has a STRESS index of 43.93 on the COM data of experimentally determined color differences. The CIE94 color difference formula has a STRESS index of 32.1 and the best performing color difference formula (CIEDE2000) has a STRESS index of 27.49 on the COM data on which it was also optimized \cite{LissnerUrban2010b}.

\section{Adjusting \A to the print size}
\label{sec:Resizing}

In many situations, it is undesirable or even impossible to 3D print an object in its original size. Scaling the object up or down has implications for its perceived translucency, however. For example, if we consider a head or full-body scan of a human, given the build space capacities of state-of-the-art 3D printers, one would most likely print it smaller than the original size--a factor of 10 is common for retail full-body ``mini-me" figurines. Using \A corresponding to the correct attenuation coefficients of human skin in this case would result in a print that looks far too translucent.

\begin{figure}[tbh]
\centering
\includegraphics[width=0.45\textwidth]{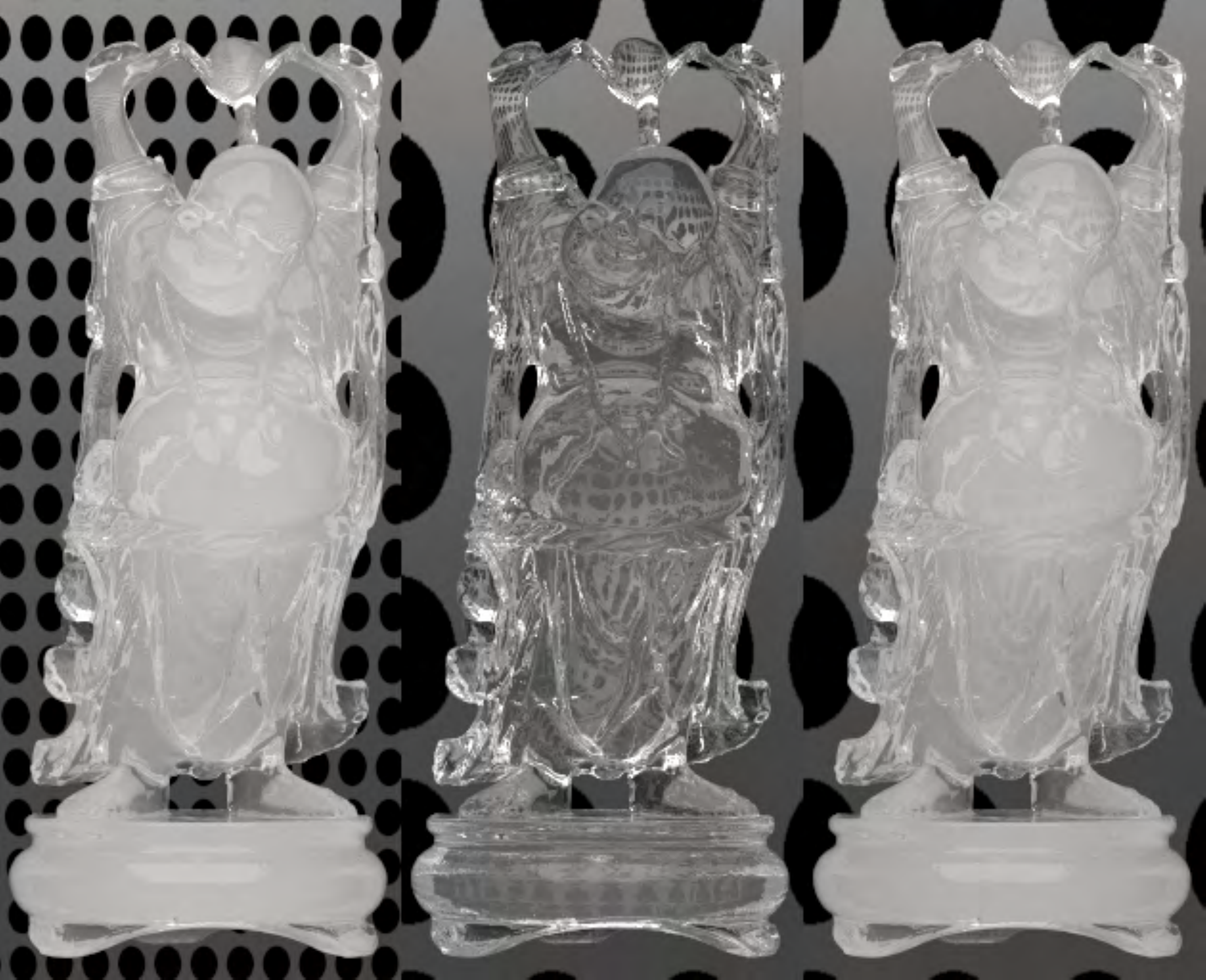} 
  \caption{Effect of object size on perceived translucency. A Buddha $123\cm$ tall (left). Scaled Buddha to $15.8\cm$ with the same (middle) and adjusted (right) \A value. For parameter values refer to Table~\ref{Table::Models}.}
	\label{Figure::Resizing}
\end{figure}

Fortunately, our definition of \A allows a simple adjustment so that resized models made of reference materials have perceptually similar translucency, as shown in Figure \ref{Figure::Resizing}. For an object with homogeneous optical material properties, the amount of light exiting the surface of the object is governed by the extinction or attenuation coefficient $\atten = \absorp + \scatt$. The \emph{mean free path}, or average distance between scattering events is given by $\meanFreeP = 1/\atten$. The optical thickness or depth $\opticalDepth$ of an object is the number of mean free paths photons travel within it before being absorbed or exiting. By keeping \A fixed we keep $\meanFreeP$ the same. If we scale the object by a factor $k$, the lengths of paths light travels will scale by $k$, the number of scattering events will scale by $k$, and $\opticalDepth$ will change accordingly. If we adjust \A so that, according to \eqref{Eq::A2ReferenceMaterials}, the \revision{scattering and absorption coefficients ($\absorp,\scatt)$} are scaled by $1/k$, we will get the same optical thickness $\opticalDepth$ as the original. The following formula accomplishes this
\begin{equation}
\A((\absorp,\scatt)/k) = (1 - (1-\A(\absorp,\scatt)^{1/q})^{1/k})^q
\end{equation}
which is derived simply by plugging $(\absorp,\scatt)/k$ into \eqref{Eq::A2ReferenceMaterials}. 

Note that in Figure \ref{Figure::Resizing}, the dot-texture in the background is kept at a fixed scale, so that the dots stay the same size, to clearly show the change in scale of the Buddha. This affects the appearance of the Buddha as the dot pattern becomes lower-frequency for the scaled-down rendering. The higher-frequency dot pattern in the larger-scale rendering is blurred by the light transport through the Buddha, resulting in a more uniform appearance--similarly to how a halftone pattern looks uniform to the human eye.

\begin{figure}[tbh]
\centering
\includegraphics[width=0.48\textwidth]{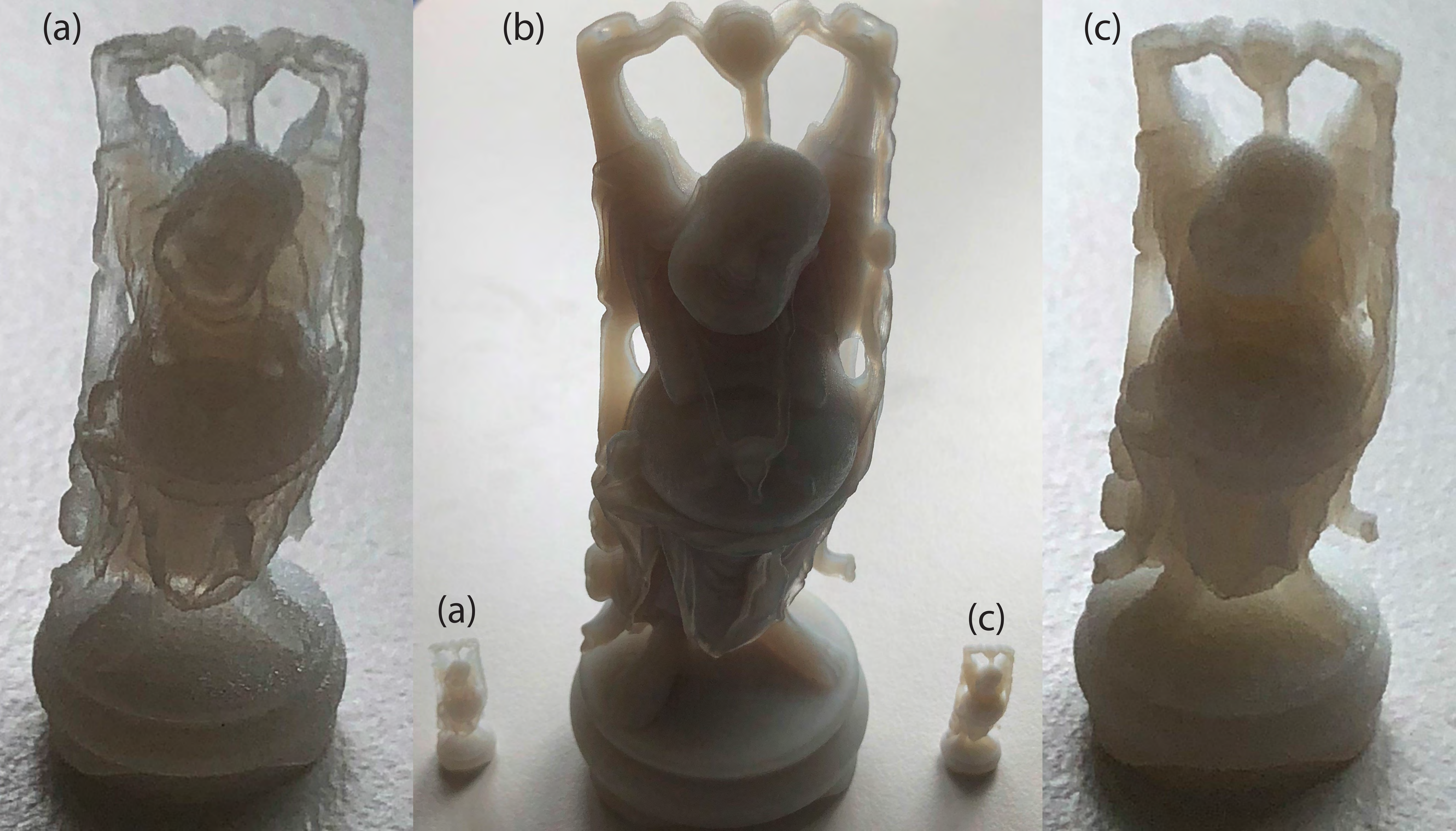} 
\caption{3D printed Buddhas scaled to (a): $3\cm$ with $\A = 0.3$, (b): $15\cm$ with $\A = 0.3$, and (c): $3\cm$ with \A scaled to match (b). Pictures were taken under backlit conditions for better comparison.}
\label{Figure::PrintedBuddhas}
\end{figure}

In Figure \ref{Figure::PrintedBuddhas} we see a similar experiment, but the models printed using a recently proposed joint color and translucency pipeline~\cite{BruntonArikanTanksaleUrban2018}. In (b), we see a $15\cm$ print with $\A=0.3$, while (a) shows a $3\cm$ print with the same \A and (c) shows a $3\cm$ print with \A scaled to match (b). Note that at the smaller size, surface roughness is more apparent in (c) compared to (b), altering the perception of translucency. However, considering the dramatic scale change, the adjusted \A matches the translucency of (b) quite well. Specific algorithmic choices of the pipeline may also have an influence, see Brunton \etal ~\shortcite{BruntonArikanTanksaleUrban2018} for details.

\section{Approximating other materials}
\label{sec:results}

\begin{table}
\centering
\caption{Models and parameters used in our examples. Red, green and blue spectra are shown in Figure \ref{Figure::AbsorptionCoefficientRGB}; for the Temple model, the blue spectrum is clipped at 100 $\cm^{-1}$.}
\begin{tabular}{cccccc}
\parbox{0.03\textwidth}{\centering height\\ ($\cm$)} & \parbox{0.03\textwidth}{\centering phase func.} & \parbox{0.04\textwidth}{\centering $\absorp$ ($\cm^{-1}$)} & \parbox{0.04\textwidth}{\centering $\scatt$ ($\cm^{-1}$)} & A & Figure\\
\hline
\multicolumn{6}{c}{Buddha} \\
\hline
$123$ & iso. & $0$ & $4.5$ & $0.19$ & \ref{Figure::Resizing} (left) \\
$15.8$ & iso. & $0$ & $4.5$ & $0.19$ & \ref{Figure::Resizing} (middle)  \\
$15.8$ & iso. & $0$ & $35$ & $0.58$ & \ref{Figure::Resizing} (right) \\
\hline
\multicolumn{6}{c}{Dragon} \\
\hline
$86.66$ & HG $0.7$ & Red$^*$ & 4 & $0.36$ & \ref{Figure::Dragon}a \\
$86.66$ & iso. & $8.2$ & $10.10$ & $0.36$ & \ref{Figure::Dragon}b  \\
$86.66$ & iso. & $8.2$ & $10.10$ & $0.36$ & \ref{Figure::Dragon}c$^{**}$  \\
$10$ & iso. & $17.2$ & $140.2$ & $0.93$ & \ref{Figure::Dragon}d  \\
$10$ & iso. & $8.2$ & $10.10$ & $0.36$ & \ref{Figure::Dragon}e  \\
$10$ & iso. & $17.2$ & $140.2$ & $0.93$ & \ref{Figure::Dragon}f$^{**}$  \\
\hline
\multicolumn{6}{c}{Lucy} \\
\hline
$130$ & HG $-0.7$ & Green$^*$ & $12$ & $0.62$ & \ref{Figure::Lucy}a \\
$130$ & iso. & $25.4$ & $29.1$ & $0.62$ & \ref{Figure::Lucy}b \\
$15$ & iso. & $25.4$ & $29.1$ & $0.62$ & \ref{Figure::Lucy}c \\
$15$ & HG $-0.7$ & Green$^*$ & $12$ & $0.62$ & \ref{Figure::Lucy}d \\
$130$ & iso. & $25.4$ & $29.1$ & $0.62$ & \ref{Figure::Lucy}e$^{**}$ \\
$15$ & iso. & $175.5$ & $294.8$ & $0.99$ & \ref{Figure::Lucy}f$^{**}$ \\
$15$ & iso. & $175.5$ & $294.8$ & $0.99$ & \ref{Figure::Lucy}g \\
$15$ & iso. & $25.4$ & $29.1$ & $0.62$ & \ref{Figure::Lucy}h$^{**}$ \\
\hline
\multicolumn{6}{c}{Temple} \\
\hline
$130$ & HG $0.3$ & Blue$^*$ & $1$ & $0.61$ & \ref{Figure::Temple}a \\
$130$ & iso. & $81.2$ & $5.8$ & $0.61$ & \ref{Figure::Temple}b \\
$15$ & iso. & $81.2$ & $5.8$ & $0.61$ & \ref{Figure::Temple}c \\
$15$ & HG $0.3$ & Blue$^*$ & $1$ & $0.61$ & \ref{Figure::Temple}d \\
$130$ & iso. & $81.2$ & $5.8$ & $0.61$ & \ref{Figure::Temple}e$^{**}$ \\
$15$ & iso. & $709.8$ & $45.1$ & $0.99$ & \ref{Figure::Temple}f$^{**}$ \\
$15$ & iso. & $709.8$ & $45.1$ & $0.99$ & \ref{Figure::Temple}g \\
$15$ & iso. & $81.2$ & $5.8$ & $0.61$ & \ref{Figure::Temple}h$^{**}$ \\ \hline
\multicolumn{6}{l}{ $^*$see Figure \ref{Figure::AbsorptionCoefficientRGB} in Appendix \ref{sec:color_test} for spectral absorption coefficients} \\
\multicolumn{6}{l}{ $^{**}$post-process coloring (see Appendix \ref{sec:color_adjust})}
\end{tabular}
\label{Table::Models}
\end{table}

\begin{figure*}[tbh]
\centering
\includegraphics[width=0.99\textwidth]{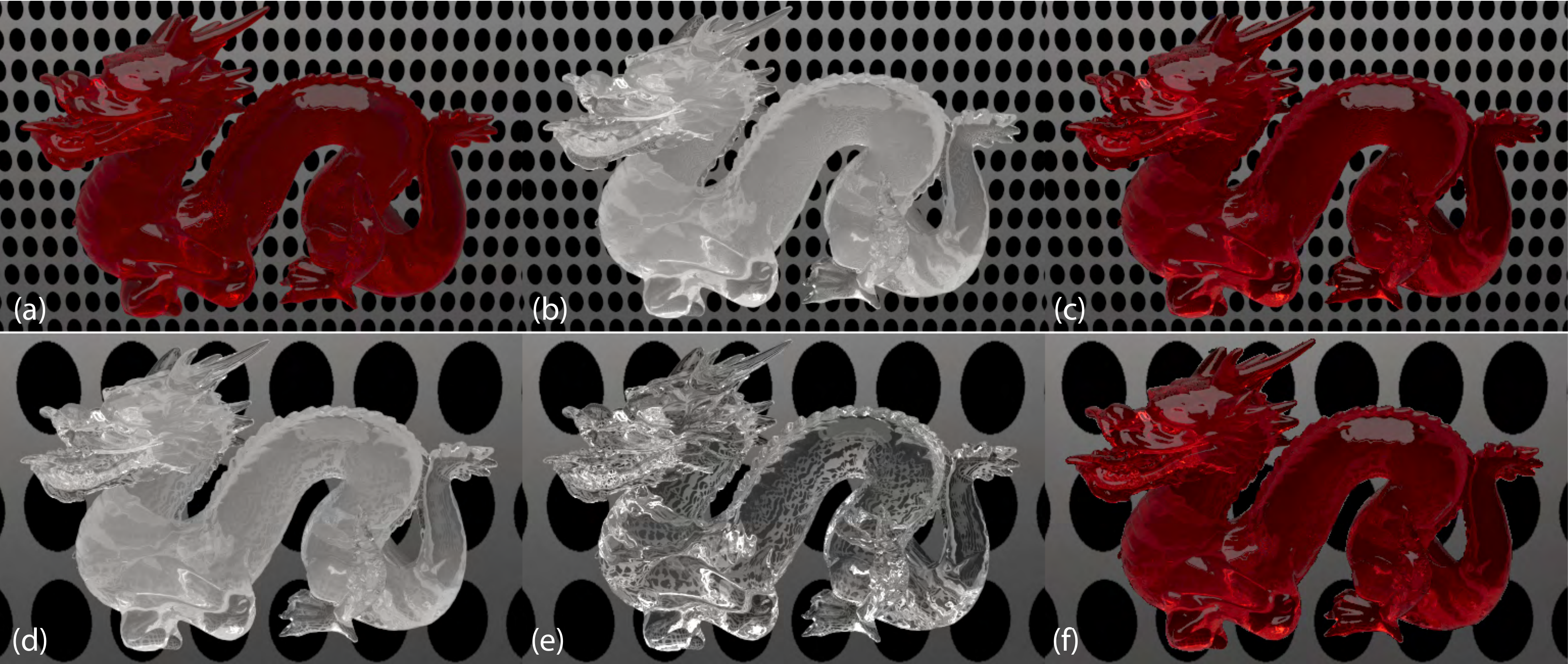}
\caption{(a): Stanford dragon rendered with a forward scattering color material. (b): a with best-fit reference material. (c): color adjustment applied to b. (d): shrunk by a factor of $8.666$ with A rescaled to perceptually match b. (e): with similar material as b. (f): color adjustment applied to d. See Table \ref{Table::Models} for details.}
\label{Figure::Dragon}
\end{figure*}

\begin{figure*}[tbh]
\centering
\includegraphics[width=\textwidth]{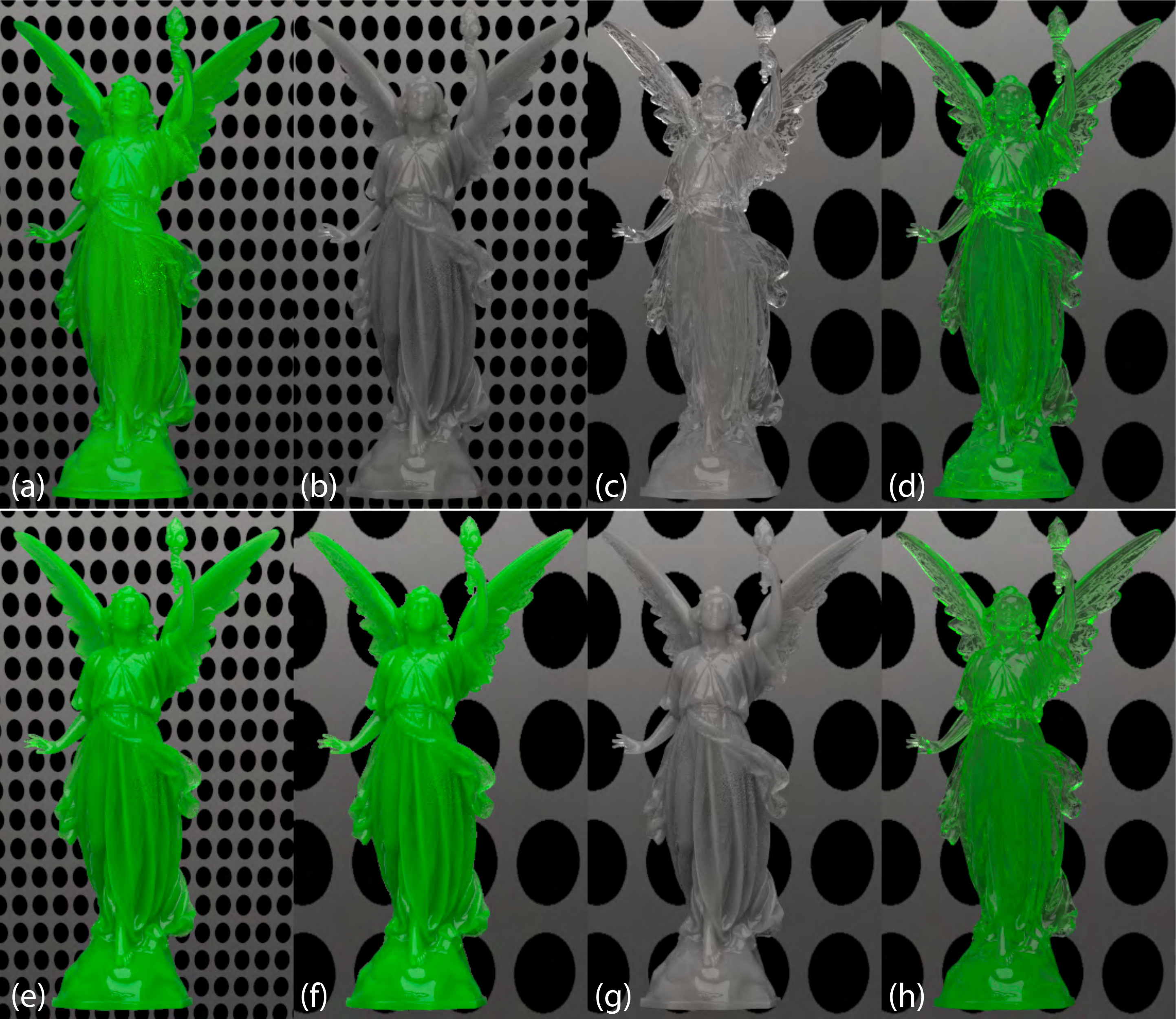}
\caption{Lucy model rendered for a backward scattering, green material scaled to (a): $130\cm$ and (d): $15\cm$ in height. (b), (c): a and d with best-fit reference material. (e), (h): color adjustment applied to b and c. (g): c with \A rescaled to perceptually match b. (f): color adjustment applied to g. See Table \ref{Table::Models} for details.}
\label{Figure::Lucy}
\end{figure*}

\begin{figure*}[tbh]
\centering
\includegraphics[width=\textwidth]{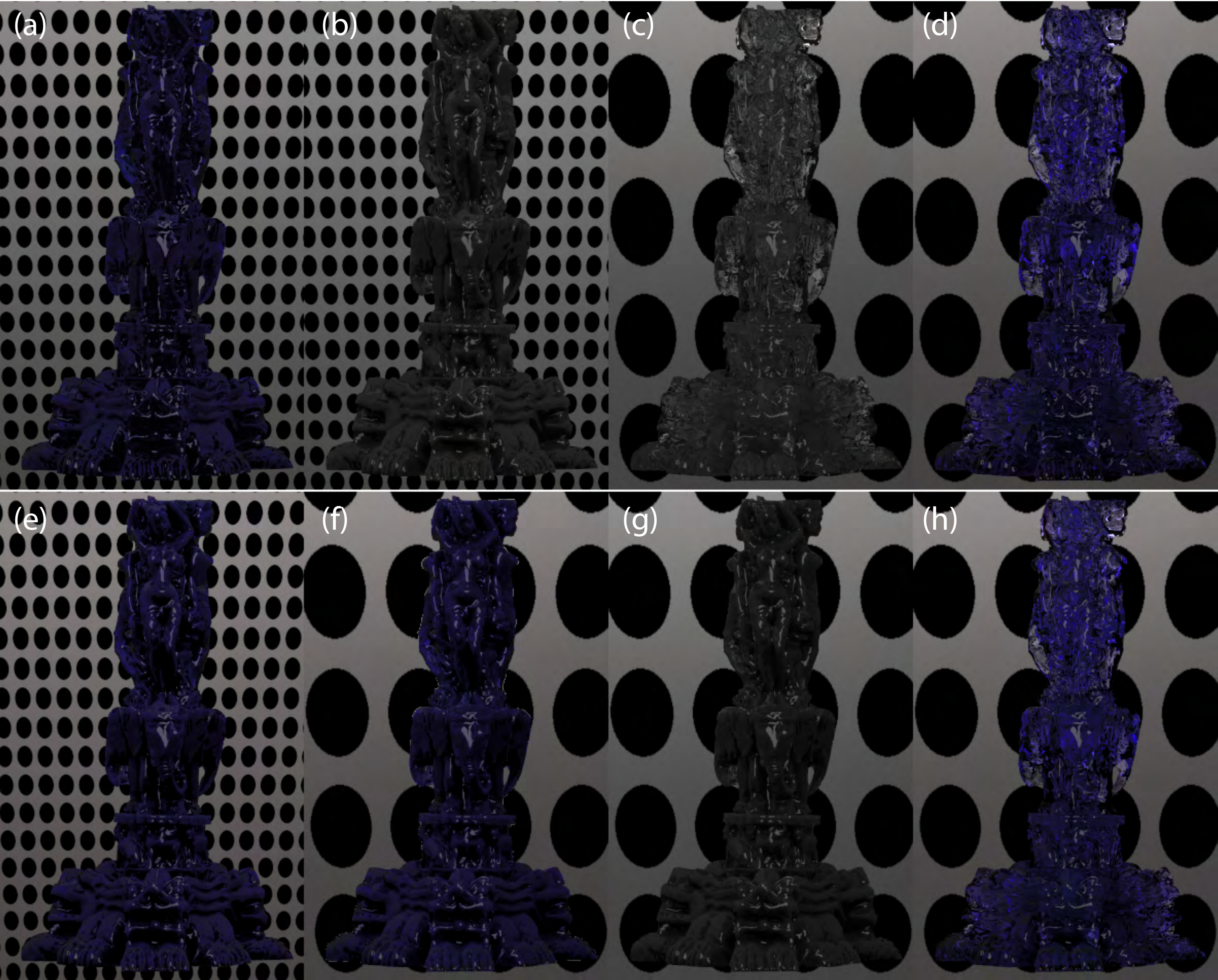} 
\caption{Temple model rendered for dark blue material scaled to (a): $130\cm$ and (d): $15\cm$ in height. (b), (c): a and d with best-fit reference material. (e), (h): color adjustment applied to b and c. (g): c with \A rescaled to perceptually match b. (f): color adjustment applied to g. See Table \ref{Table::Models} for details.}
\label{Figure::Temple}
\end{figure*}

Our definition of \A represents a small fraction of possible optical materials, specifically a 2D subset spanned by the reference materials. We must therefore consider how to treat the remaining space of optical materials.

Given a material description in the form of optical material parameters, we can find the best matching reference material by simulating the measurement setup as described in Section \ref{Subseq::Measurements} and linking the simulated measurements to the references materials as described in Section \ref{sec:DefiningA:Linking}. In the top row of Figure \ref{Figure::Dragon}, we see on the left the Stanford Dragon rendered with material properties not spanned by the reference materials, and in the middle we see the rendering with the reference material to which it is mapped. Figure \ref{Figure::Lucy} and \ref{Figure::Temple} show the same for the Lucy and Temple model. Table \ref{Table::Models} gives the parameter values for all renderings.

The measurement setup in Section \ref{Subseq::Measurements} discards color information, as the reference materials are defined to have wavelength-independent absorption coefficients. An image-space post-process transfers color from the original to the reference rendering solely to simplify the comparison, as described in Appendix \ref{sec:color_adjust}. Note that the post-process does not change lightness contrast in the non-specular areas and thus preserves the translucency cues provided by the reference materials according to Motoyoshi~\shortcite{Motoyoshi2010}.

The selected materials should illustrate the limitations of our \A definition: 
They posses strong forward (Dragon) and backward (Lucy) scattering as well as a very dark (small lightness) material that is highly absorbing for almost all wavelength for which the luminous efficiency function of the HVS is high but almost transparent for other wavelengths (Temple) (see blue curve in Figure \ref{Figure::AbsorptionCoefficientRGB}). 

Using isotropic scattering to mimic forward scattering materials causes increasing lightness contrast for front/side-lit condition as can be seen in the Dragon's leg area (compare Figure \ref{Figure::Dragon} (a) and (c)). The opposite can be seen for backward scattering where the lightness contrast is smaller for the reference materials as can be seen in Lucy's chest region (compare Figure \ref{Figure::Lucy} (d) and (h)). 

A very dark material with negligible scattering and wavelength-dependent absorption that is zero for a few wavelength shows higher apparent translucency than the best-fit reference material possessing a wavelength-independent high absorption causing incident light to be almost fully absorbed in the thick object (magnify Figures \ref{Figure::Temple}(a) and (b)/(c)). For the small Temple, light of all wavelengths is passing through the object causing luminance contrasts and thus the apparent translucency of original and reference material to be very similar (compare Figure \ref{Figure::Temple} (d) and (c)/(h)).

\subsection{\A of Real Materials}
\label{sec:real_materials}
To validate \A as a translucency space for real materials, we conducted measurements of real materials, and printed patches using the resulting sRGB and \A values using the pipeline of Brunton \etal\ ~\shortcite{BruntonArikanTanksaleUrban2018}. We chose samples of the following materials: a green wax, salmon, a green stone, a violet stone, and a green soap. Table \ref{tab:measurements} in Appendix \ref{sec:measure_real} shows the measured sRGB and \A for a set of sample materials, and the results of putting those values through the 3D printing pipeline. Note that the errors reflect a combination of double measurement error (original and printed patch), plus gamut mapping and other reproduction errors (see Brunton \etal\ for details).

\section{Limitations}
\label{sec:limitations}

Given we are consider just a one-parameter family of BSSRDFs, this comes with some limitations. By choosing an isotropic scattering function, \A is limited in its ability to approximate materials with anisotropic scattering. Figures \ref{Figure::Dragon}, \ref{Figure::Lucy} and \ref{Figure::Temple} show how well this approximation works for materials with different Heyney-Greenstein phase functions and spectral absorptions.

Measuring and assigning \A to heterogeneous materials works well if the heterogeneity is uniformly distributed, which importantly includes halftoned 3D printing materials used to characterized a printing system. However, stratified materials with a thickness exceeding the sample thickness used for measurements cannot be measured.

Our measurement setup is restricted to slab geometry, and measuring \A \emph{in the wild} is a highly challenging problem, which we leave for future work.

Further, we propose \A as a translucency \emph{space}, not a translucency \emph{appearance model}, which has a number of consequences. The impact of other visual attributes, so called cross-contamination of visual attributes, is not considered, meaning the effect of gloss on perceived translucency is not modeled. The visual appearance of a material with a given \A is dependent on the defined viewing conditions (Sec. \ref{sec:PsychophysicalExperiment:viewcond}), and deviations from those conditions will affect the perceptual uniformity of $\A$.

\section{Conclusion}
\label{sec:Conclusion}

We have presented a new interpretation, or redefinition, of the \A channel in RGBA, designed for graphical 3D printing, which replaces the traditional additive blending interpretation with a subtractive mixing one for material translucency. Our interpretation links \A to physical material measurements and embeds \A in a nearly perceptually uniform translucency scale for a set of virtual reference materials. By linking \A to these reference materials, we maintain device independence for both measurements and reproduction. Our definition allows simple adjustment to maintain consistent translucent appearance in the presence of object scaling, which is common in graphical 3D printing. 

Our interpretation only considers isotropic phase functions, and therefore incorporating or adapting the interpretation to anisotropic phase functions of materials found in our environment is an important aspect of future work. Related to this is an investigation of the number of bits needed, from perceptual point of view, to sample $\A$, as defined in this paper, and how many bits are needed to either represent commonly used phase functions or to enumerate them. One option we consider promising is the notion of an $\A$-\emph{context}, which would encompass additional parameters influencing the perception of translucency, such as phase function representations. Different contexts (phase functions) would give rise to different reference materials, and could themselves be standardized. We expect that printing materials for specific devices will mostly differ in terms of scattering and absorption, \ie~$\A$, rather than phase functions, which means contexts could be assigned to entire objects with spatially varying $\A$. 

Other key direction for future work include further exploration of the link between perception of color and perception of translucency, and techniques to obtain \A for materials for which slab geometry is difficult to obtain.

\begin{acks}
We thank the anonymous reviewers for their insightful and constructive comments. We thank Markus Barbieri for providing the design and Martin Majewski for creating the CAD data to simulate the Barbieri Spectro LFP, all observers who participated in our psychophysical experiment, and Can Ates Arikan for helpful discussions. This work was funded by the FhG Internal Programs under Grant No. Attract 008-600075, Scan4Reco project funded by EU Horizon 2020 Framework Programme under grant agreement no 665091 and AIF IGF-Vorhaben Nr. 18478 as well as JSPS KAKENHI Grant Number JP15H05922, JP16J00273 and the Leading Graduate School Program R03 of MEXT.
\end{acks}

\bibliographystyle{ACM-Reference-Format}

\pagebreak
\appendix

\section{Meaningful edge-loss measurements}
\label{sec:meaningful}

In order to be meaningful, the edge-loss setup must measure lateral light transport effects visible for the human observer for a typical viewing distance. In other words: The low-pass filtering effect caused by lateral light transport detectable by the setup must be at spatial frequencies for which the human contrast sensitivity is not zero. To validate this, we used a set of isotropic point spread functions (PSF) $\vec{p}(x,c) = \exp(-c\|x\|_2^2)$, whose parameters $c$ are chosen so that for each integer resolution within the visible resolution range of [0, 50] cycles/degree (cpd) at least one PSF shows a decay of 50\% assuming a viewing distance of $80\cm$\footnote{Gaussian-type PSFs were selected to create a range of blurring magnitudes without aiming to mimic PSFs of distinct materials.}. For higher resolutions the achromatic contrast sensitivity is almost zero for office luminance conditions \cite{Barten1999}. We simulated the measurement $m(a_i,c)$ of a material with a PSF parametrized by $c$ with an aperture $a_i$ for illumination as follows 
\begin{eqnarray}
m(a_i,c) = \frac{1}{\pi}\int_{\vec{D}} \vec{circ}(x,a_i) * \vec{p}(x,c) \ dx
\end{eqnarray}
where $\vec{D} = \{x \in \mathbb{R}^2 | \|x\|_2 \leq a_d/2 = 1\}$ is the detection area, $*$ is the convolution operator and the irradiance is modeled by $\vec{circ}(x,a_i) = 1$ for $\|x\|_2 \leq a_i/2$ and 0 elsewhere. This is a simplification because the irradiance within the illuminated area is generally not uniform for real spectrophotometers. 

\begin{figure}[tb]
\centering
\includegraphics[width=0.45\textwidth]{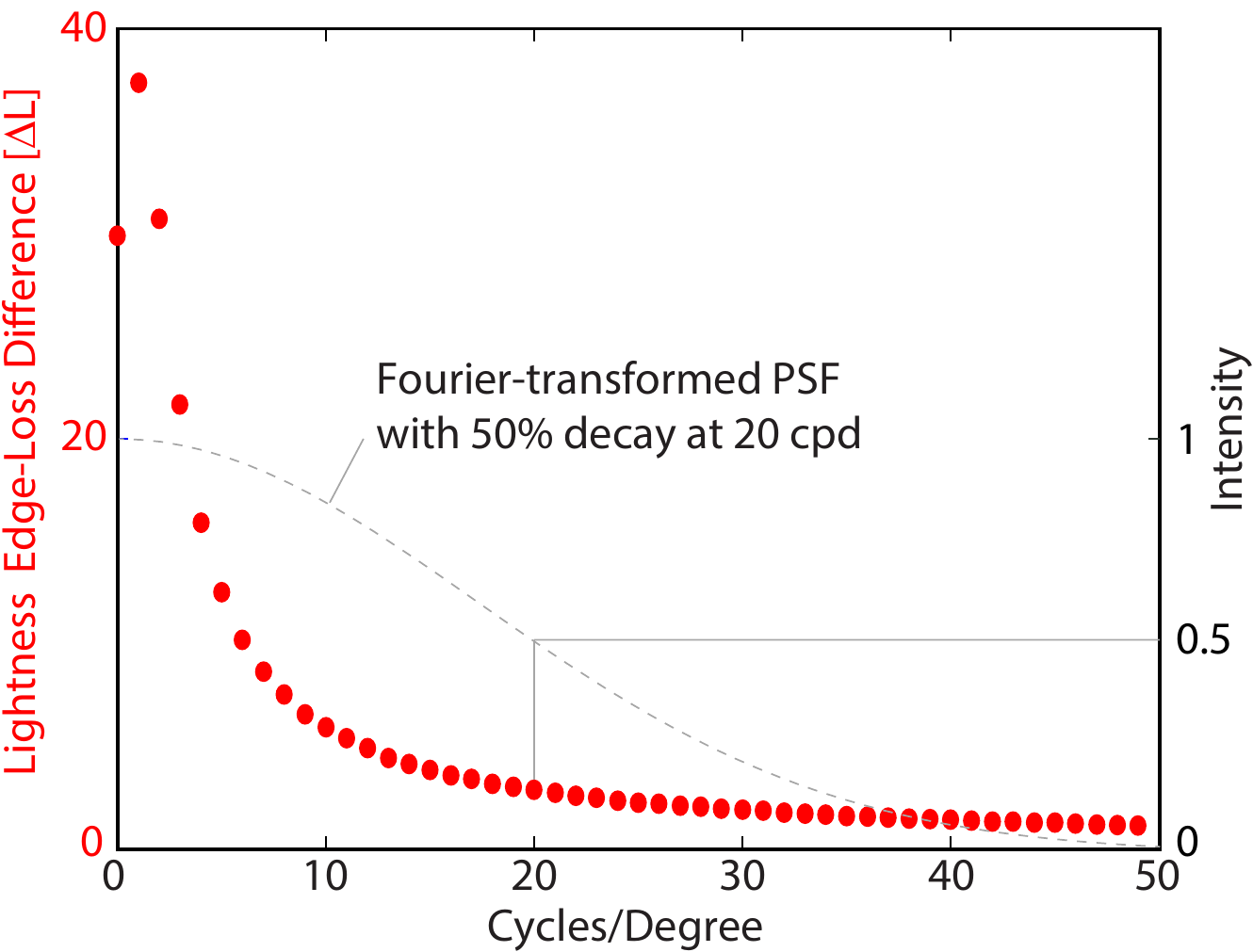} 
  \caption[width=\textwidth]{Each red dot indicates the edges-loss measurement for a PSF with a decay of 50\% at the corresponding cpd.}
	\label{Figure::ContrastSensitivity}
\end{figure}

Figure \ref{Figure::ContrastSensitivity} shows the edges-loss measurements for conditions $a_0 =~1$ and $a_1= 1/4$ with $a_d = 2 \mm$. All edges-loss values are larger than $\Delta L^* = 1$ and therefore detectable by spectrophotometers used in graphic arts. 

Detection of these PSFs is one of two necessary criteria of the edge-loss measurement setup to be meaningful. The other is the ability to discriminate the blurring effect caused by these PSFs, which must be comparable or better than that of the HVS. The discrimination is limited by the repeatability error of the spectrophotometric measurments employing the two apertures. This repeatability error results in a measurement uncertainty that makes it impossible to discriminate PSFs with very similar edge-loss measurements; see particularly measurements for large cpd values in Figure \ref{Figure::ContrastSensitivity}. To discriminate two edge-loss measurements, we used a spectrophotometric discrimination threshold of $0.5$ based on average short-term repeatability errors of handheld spectrophotometer of $\Delta E^*_{ab} = 0.12$ and medium-term repeatability of $\Delta E^*_{ab} = 0.24$ \cite{WybleRich2007a}. 

\begin{figure}[tbh]
\centering
\includegraphics[width=0.45\textwidth]{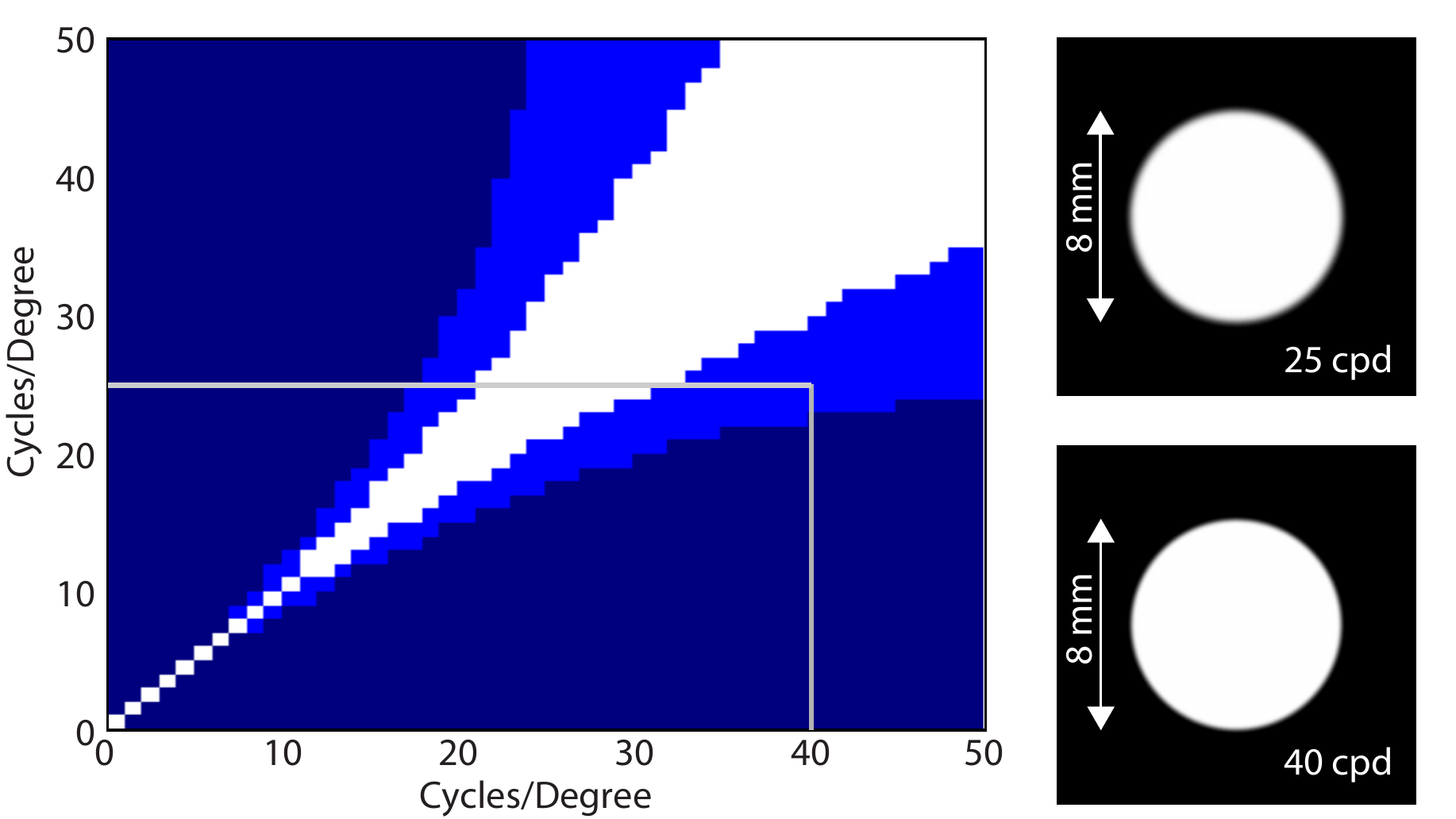} 
  \caption[width=\textwidth]{Comparison between the abilities of the HVS and the spectrophotometer to discrimination the blurring effect described by the investigated PSFs: The HVS and the device can both discriminate PSFs (dark blue); only the device can discriminate PSFs (blue); neither the HVS nor the device can discriminate PSFs (white). The PSFs with a decay of 50\% at 25 cpd and 40 cpd computed for a viewing distance of $80 \cm$ are exemplary marked in the diagram. On the right side the blurring effect caused by these PSFs is shown for a disk-shaped uniformly illuminated area with a diameter of $8\mm$. The blurring effect caused by these PSFs can be distinguished by the spectrophotometer but not by the HVS from a viewing distance of $80 \cm$.}
	\label{Figure::ContrastSensitivityDiscrimination}
\end{figure}

For estimating the ability of the HVS to discriminate two PSFs, we compared images blurred by these PSFs using the approach described by Johnson and Fairchild~\cite{JohnsonFairchild2003}. For a survey of other models of blur discrimination we refer to \cite{WatsonAhumada2011}. 
As a test image containing a sharp edge with maximum contrast, we used the image described by $\vec{circ}(x,8 \mm)$, which models a disk-shaped uniformly illuminated area with a diameter of $8\mm$. For modeling light transport, we convoluted the image with the PSF (see Figure \ref{Figure::ContrastSensitivityDiscrimination}, right) and for modeling the contribution of the viewing distance, we convoluted the blurred image with the achromatic \emph{Contrast Sensitivity Function} (CSF). We used the CSF employed by the iCAM framework for evaluating image differences ~\cite{FairchildJohnson2004}. Each of the 50 considered PSFs corresponds to an image blurred by light transport and the CSF. For comparing two PSFs, we computed pixel-wise lightness differences from the corresponding images and took the average $\Delta L^{*}_{\textrm{mean}}$ employing only non-zero values. We assumed that for $\Delta L^{*}_{\textrm{mean}} < 1$ the blurring effects caused by the corresponding PSFs are not distinguishable by the HVS. Note that the perception of lightness and color differences depends on various factors, such as the luminance, sample size, texture, or background color~\cite{CIEParametric1993}. For $120 \unit{cd/m}^2$ (the luminance of typical LCD displays), $\Delta L^{*}_{\textrm{mean}} = 1$ is below the just noticeable difference (JND)~\cite{UrbanFedutinaLissner2011}. Since the samples are non-uniform, the selected threshold is likely much lower than the real one~\cite{MontagBerns2000}, i.e. the ability of the HVS to discriminate lightness variations is even lower that assumed here. 

Figure \ref{Figure::ContrastSensitivityDiscrimination} compares the ability of the HVS and the spectrophotometric edge-loss setup to discriminate the evaluated PSFs. It shows that the instrument is either equally or more sensitive than the HVS. This statement holds even if we would increase the spectrophotometric discrimination threshold to $0.8$. In summary, we can conclude that the edge-loss setup to measure lateral light transport employing common spectrophotometers used in graphic arts is meaningful with respect to typical viewing distances of approx. $80 \cm$. 

\section{Standardized residual sum of squares (STRESS) index}
\label{Appendix:STRESS}
The STRESS index is used as a performance measure in multidimensional scaling (MDS) and to evaluate the performance of color difference equations \cite{MelgosaHuertasBerns2008}. It is defined in our context as follows
\begin{eqnarray}
\textrm{STRESS} & = & 100 \left(\frac{\sum_{i=1}^{n}\left( \Delta T_{i} - G\Delta V_i \right)^2}{\sum_{i=1}^{n} G^2\Delta V_i^2}\right)^{1/2} \\
G & = & \frac{\sum_{i=1}^{n} \Delta T_i^2}{\sum_{i=1}^{n} \Delta T_i \Delta V_i} \nonumber
\end{eqnarray}
where $\Delta T_{i}$ are computed and $\Delta V_i$ are visual translucency differences of a sample pair $i$ in a set of $n$ investigated pairs. The STRESS index is always in the range of [0,100], where 0 indicates a perfect agreement between computed and visual translucency differences. The larger the STRESS index the larger is the disagreement. The STRESS index allows a simple significance comparison between two formulas, $X$ and $Y$, computing translucency differences by comparing the term 
\begin{eqnarray}
F = \left(\frac{\textrm{STRESS}_X}{\textrm{STRESS}_Y}\right)^2
\end{eqnarray}
with the critical value $F_C$ of the two-tailed $F$-distribution with 95\% confidence level and $(n-1,n-1)$ degrees of freedom: if $F < F_C$, formula $X$ is significantly better than $Y$, if $F > 1/F_C$, formula $X$ is significantly poorer than $Y$, otherwise both formulas perform insignificantly different.

\section{Color Test Pairs}
\label{sec:color_test}

\begin{figure}
\centering
\includegraphics[width = 0.45\textwidth]{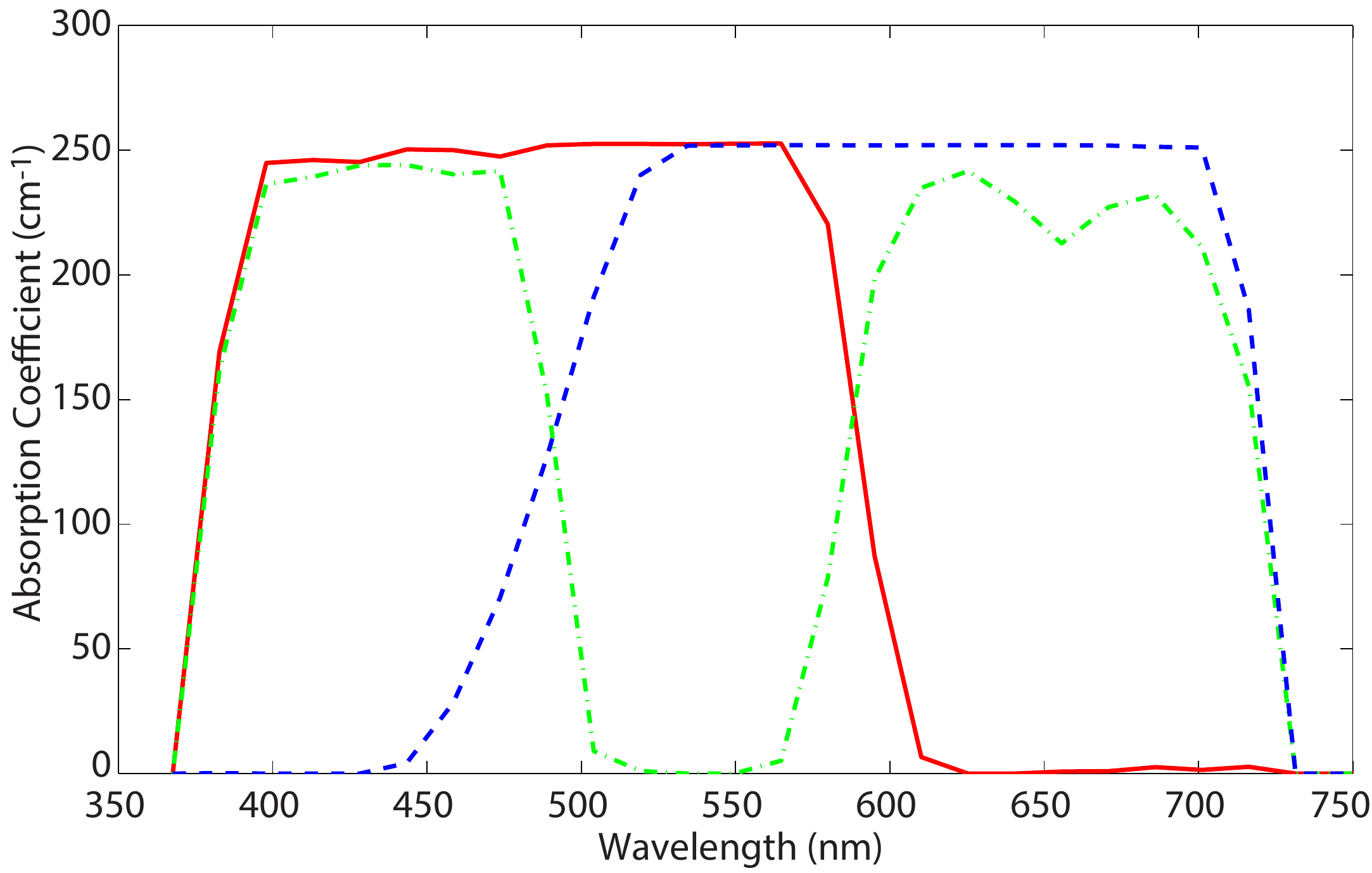} 
  \caption{Absorption coefficients used for the second set of test pairs in the visual experiment}
	\label{Figure::AbsorptionCoefficientRGB}
\end{figure}

\begin{figure}[ht!]
\centering
\includegraphics[width=0.48\textwidth]{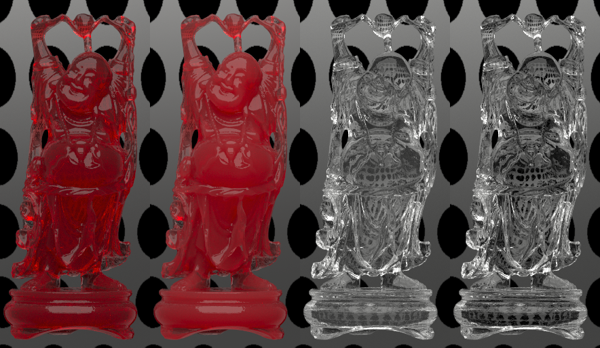} 
  \caption{Example images shown to subjects: Red center/test pair (left) and anchor pair (right)}
	\label{fig:BuddhaExptExampleRed}
\end{figure}

Figure \ref{Figure::AbsorptionCoefficientRGB} shows the spectra used for the second set of pairs with non-zero absorption, resulting translucent color test pairs.
Figure \ref{fig:BuddhaExptExampleRed} shows example images shown to the subjects for the red spectra.

\section{Comparing Perceptual-Uniformity of \A and \Ahat}
\label{sec:PerceptuallyUniformityAAhat}
\begin{figure*}[tbh]
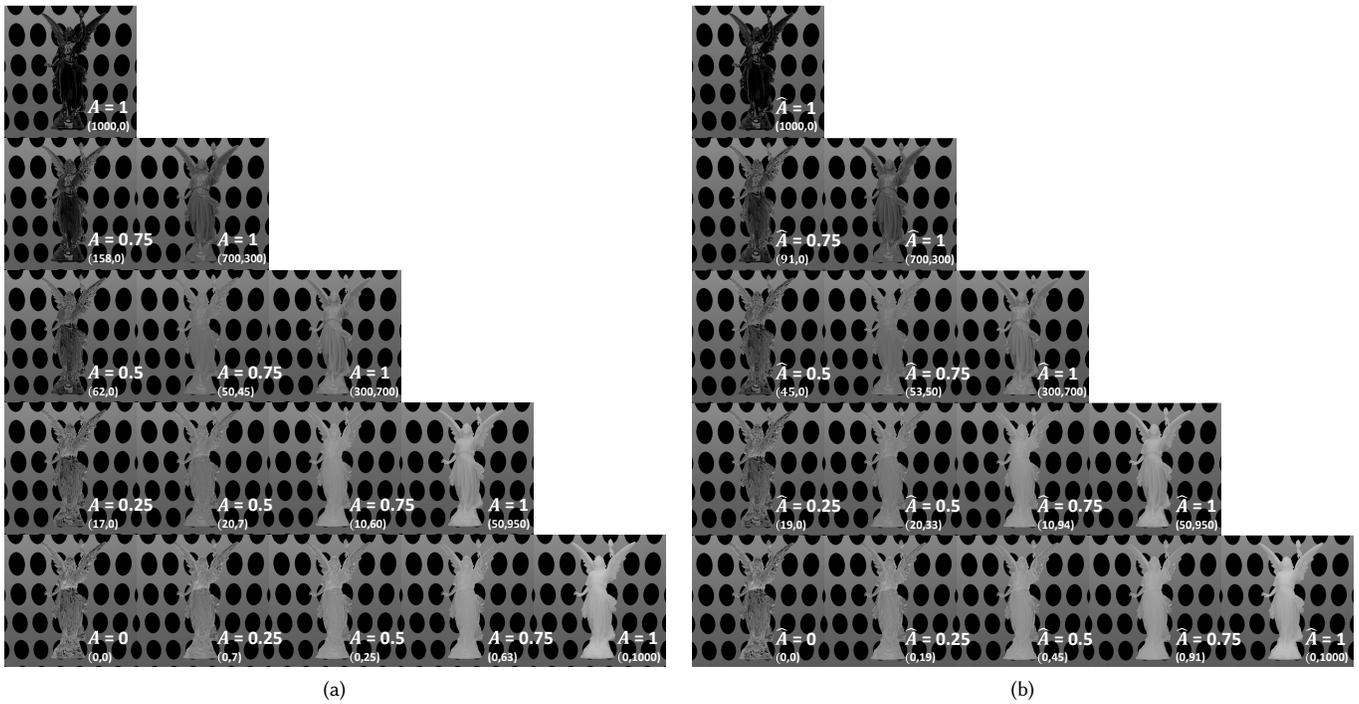

    \centering
    \subfloat[]{\includegraphics[width = 0.49\textwidth]{Uniform_A.pdf}}
		\hspace*{\fill}
		\subfloat[]{\includegraphics[width = 0.49\textwidth]{Uniform_Ahat.pdf}}
    \caption{Lucy-renderings corresponding to a uniform sampling of $\A$ (a) and $\Ahat$ (b) for reference materials covering the visually-relevant absorption and scattering range. The numbers in the brackets correspond to $(\absorp, \scatt)$. $\Ahat$ is defined similarly as $\A$ but with a modified attenuation coefficient and a psychometric function parametrized with $(p,q) = (1,1)$.}
		\label{Figure::AAhat}
\end{figure*}

\revision{Figure \ref{Figure::AAhat} demonstrates the perceptual uniformity of $\A$ compared to that of $\Ahat$. Figure \ref{Figure::AAhat}a shows renderings of the Lucy model with scattering and absorption parameters corresponding to a uniform $\A$-sampling of the reference materials covering the visually-relevant absorption and scattering range, whereas Figure \ref{Figure::AAhat}b shows the same for a uniform sampling in $\Ahat$. The improvement of perceptual uniformity of $\A$ compared to $\Ahat$ is particularly apparent in scattering ramps. Best view digitally by zooming in.} 

\revision{Since multiple reference materials correspond to the same $\A$- or $\Ahat$-value, these renderings illustrate just one of many possible selections of reference materials. We selected  for identical $\A$- or $\Ahat$-values (diagonals in each subfigure) reference materials sampling the possible lightness reflectance range equidistantly.}

\section{Measurments of Real Materials}
\label{sec:measure_real}

Table \ref{tab:measurements} shows the sRGB and \A values measured for some real materials, and the errors of printed patches specified with those values using the pipeline of Brunton \etal\ ~\shortcite{BruntonArikanTanksaleUrban2018}. The samples of violet stone, green stone and green soap can be seen in Figure \ref{fig:lucy_materials} (right), and the results of printing with the values for green wax can be seen in Figure \ref{fig:wax_ships} (second from right). 

\begin{table}[ht]
\centering
\caption{Measurements of real materials and errors of patches printed with the same values.}
\begin{tabular}{c|cc|c|c}
Material			& \multicolumn{2}{c|}{Measured from original} & \multicolumn{2}{c}{Errors of printed patch} \\
							& sRGB 						& A			&	CIEDE2000 & dA	\\
\hline
green wax			&	[0, 0.30, 0.27]			&	0.6		&	3.4471		& 0.1323\\
salmon				&	[0.89, 0.56, 0.33]	&	0.72	&	6.1587		& 0.0413\\
green stone		&	[0.39, 0.40, 0.20]	&	0.49	&	2.8690		& 0.2509\\
violet stone	&	[0.23, 0.05, 0.26]	&	0.68	&	3.3113		& 0.0460\\
green soap		&	[0.77, 0.82, 0.69]	&	0.157	&	8.1293		& 0.0019
\end{tabular}
\label{tab:measurements}
\end{table}

\section{Additional 3D Printed Examples}
\label{sec:additional_prints}

\begin{figure}[ht]
\centering
\includegraphics[height=8.0cm]{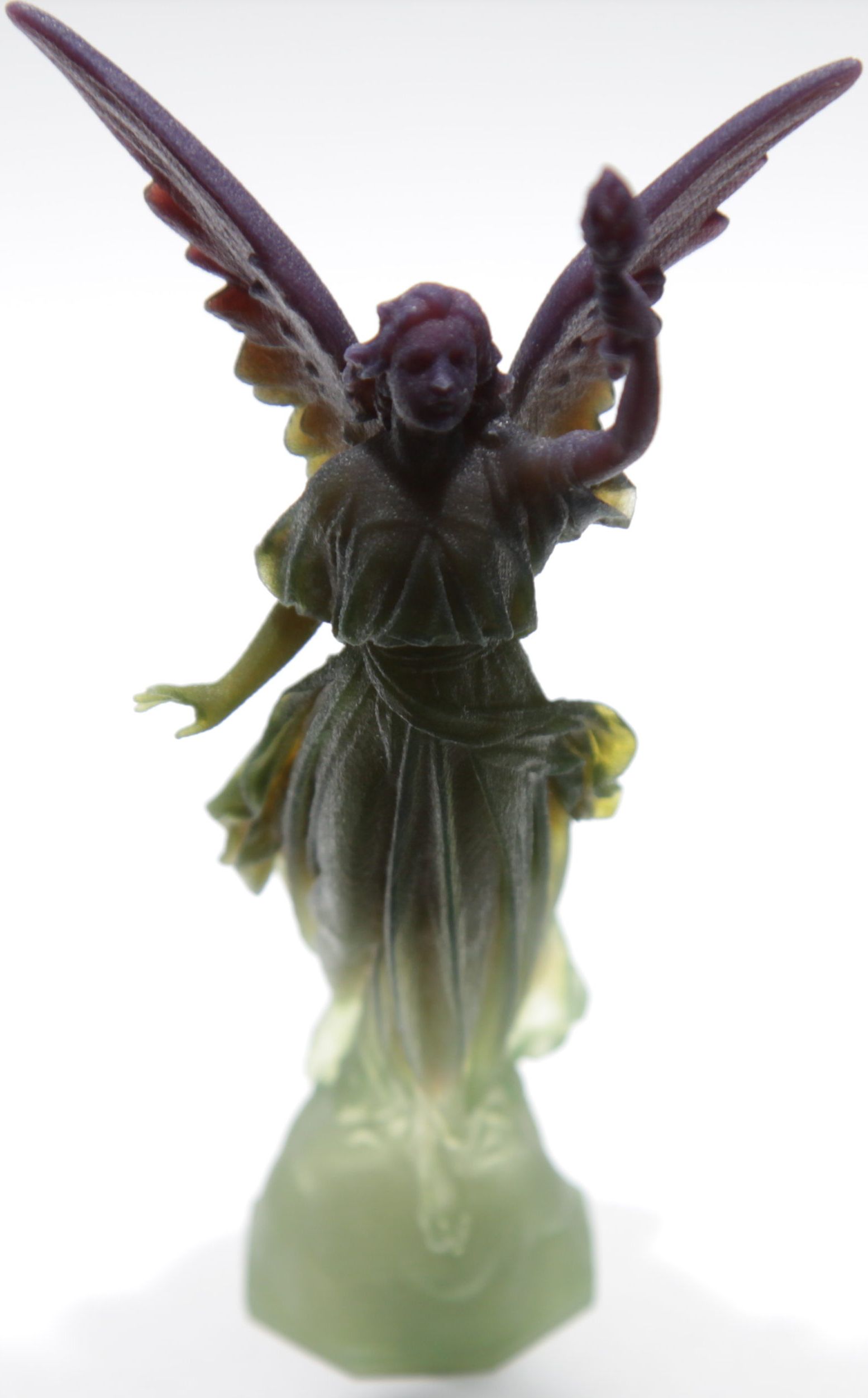}
\hfill
	\parbox[b][8.0cm][s]{0.19\textwidth}{
	\centering
	\includegraphics[height=2cm]{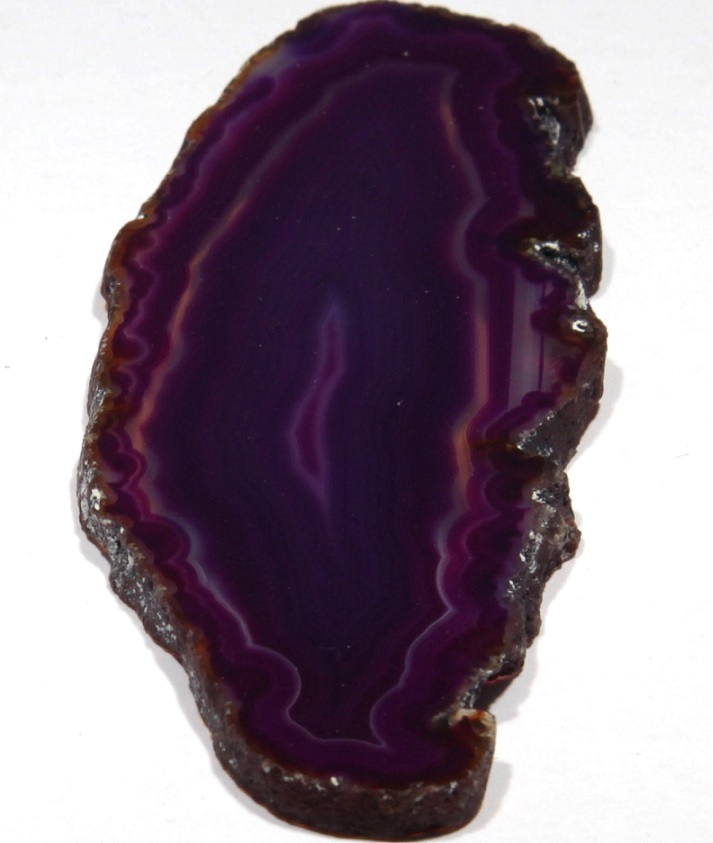}\\
	\vfill
	\includegraphics[height=2cm]{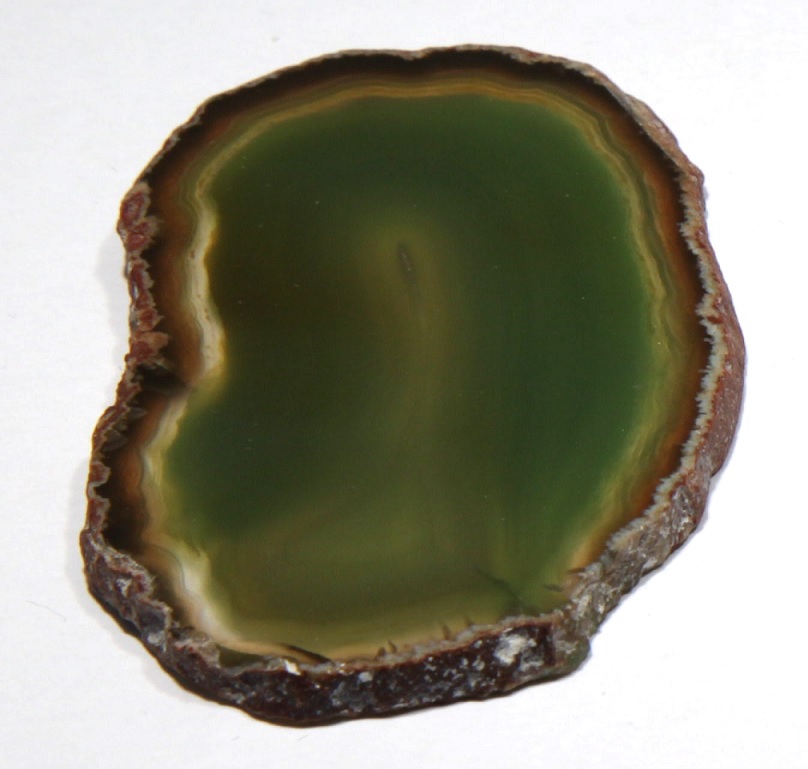}\\
	\vfill
	\includegraphics[height=2cm]{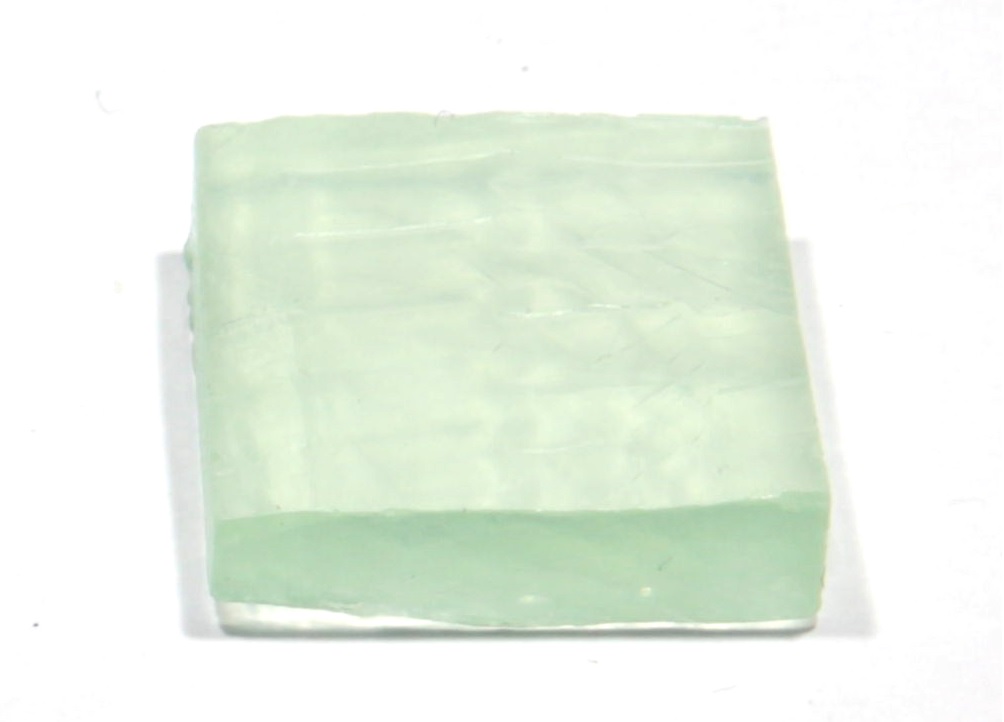}\\
	}
\caption{The St. Lucy model printed ($10\cm$) with varying RGBA values. At the top is that of the violet stone, middle is green stone and bottom is green soap, with linear transitions in between. Figure reproduced from Brunton \etal~\protect\shortcite{BruntonArikanTanksaleUrban2018}.}
\label{fig:lucy_materials}
\end{figure}

\begin{figure}[!ht]
\includegraphics[width=0.23\textwidth]{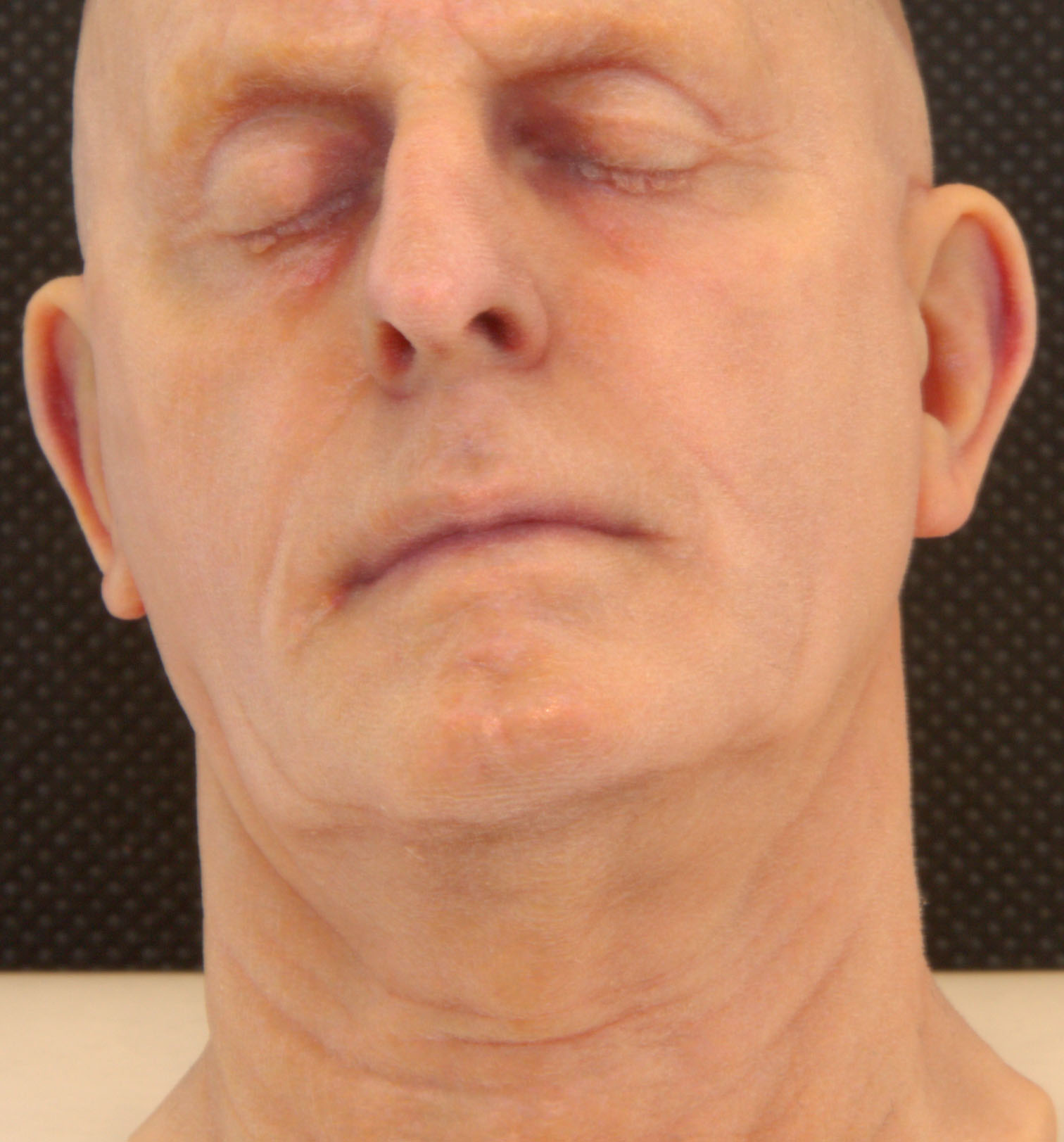}
\hfill
\includegraphics[width=0.23\textwidth]{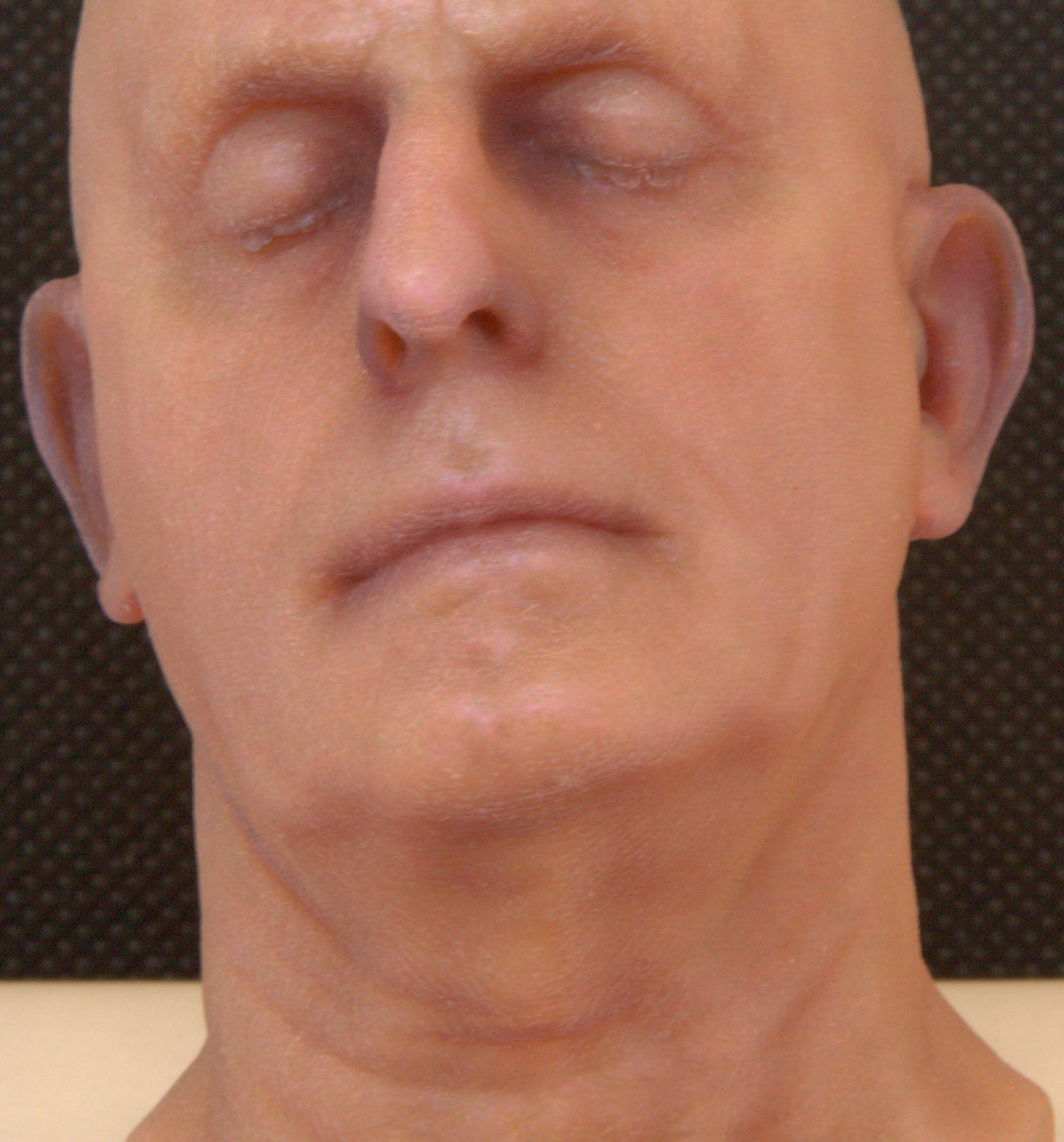}
\caption{Head model printed with $\A=0.786$ (left) and $\A=0.518$ (right). Identical model geometry and illumination conditions. Figure reproduced from Brunton \etal~\protect\shortcite{BruntonArikanTanksaleUrban2018}.}
\label{fig:HeadOpaqueTranslucent}
\end{figure}

\begin{figure*}[!ht]
\includegraphics[width=0.245\textwidth]{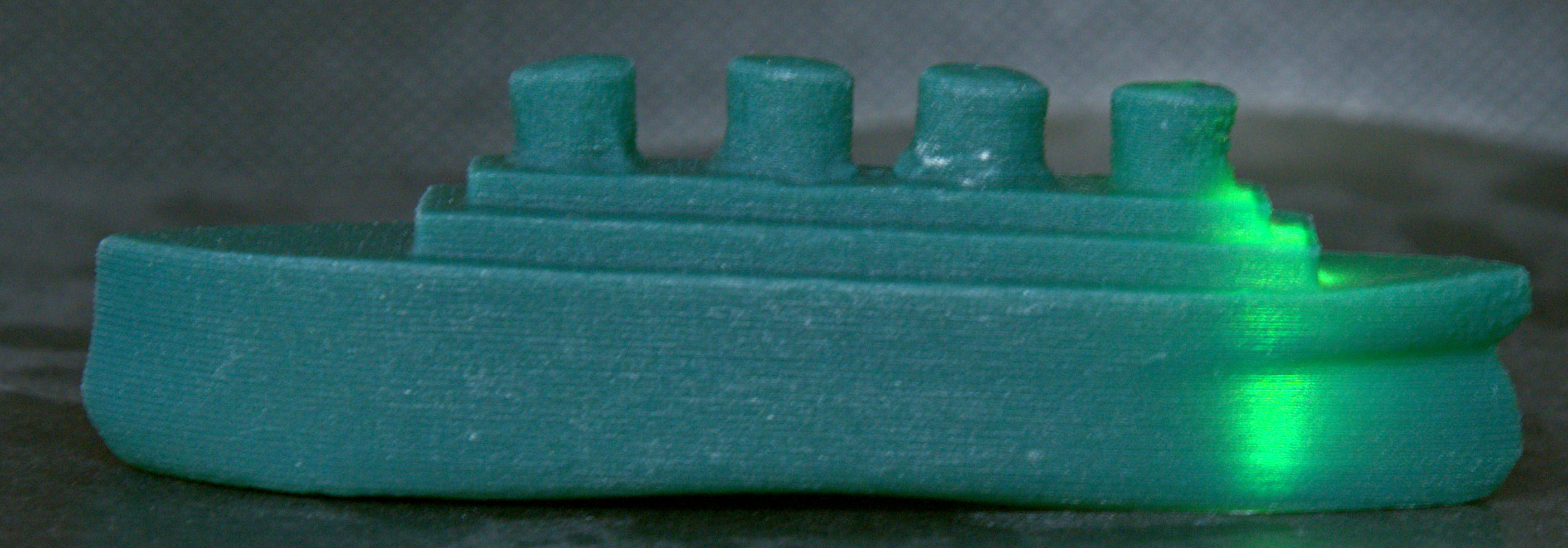}
\hfill
\includegraphics[width=0.245\textwidth]{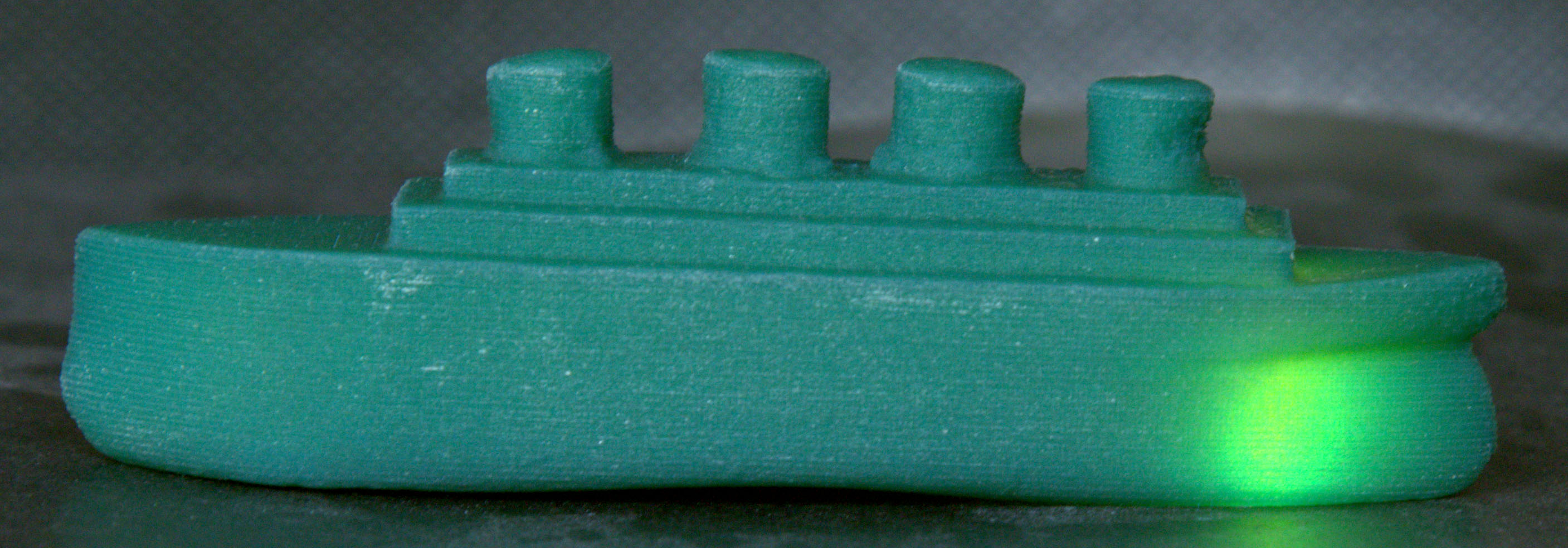}
\hfill
\includegraphics[width=0.245\textwidth]{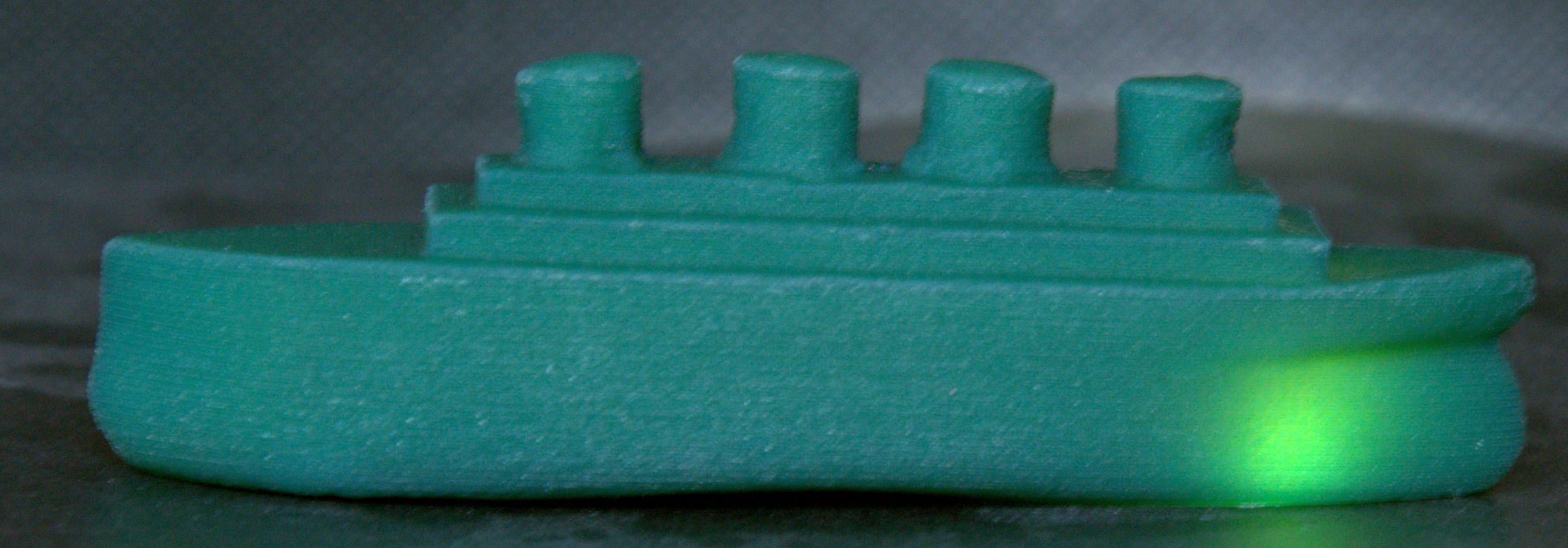}
\hfill                    
\includegraphics[width=0.245\textwidth]{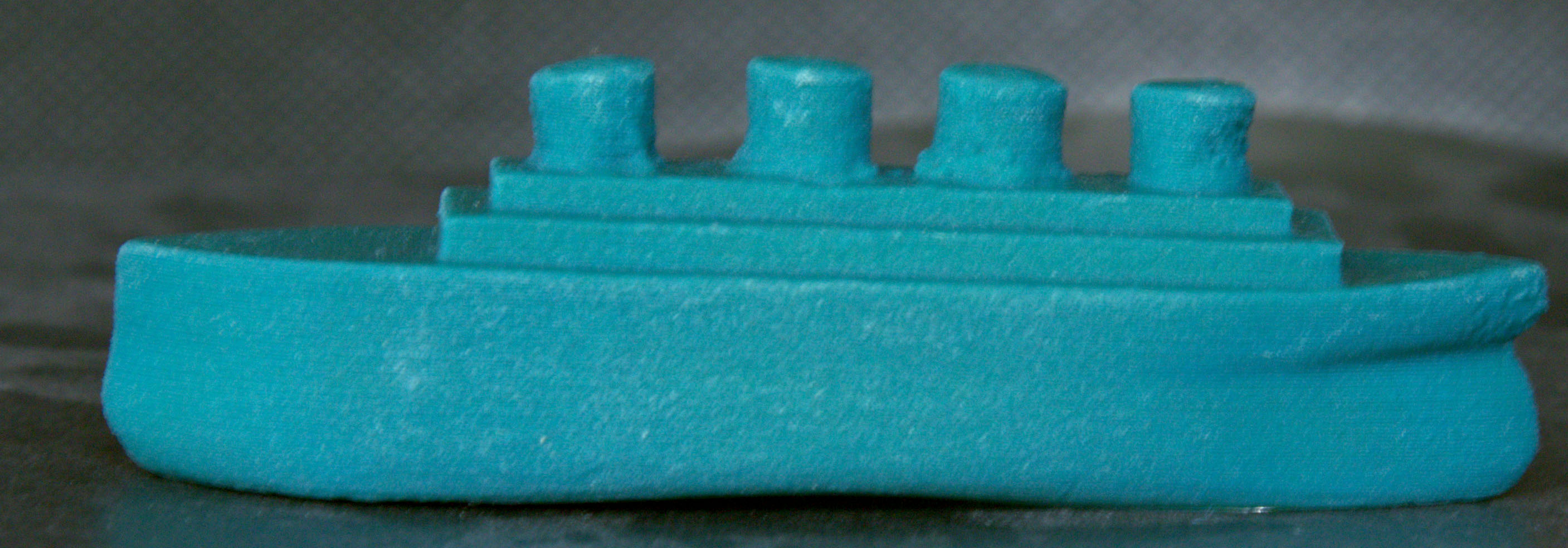}
\caption{3D prints with color measured from a sample of wax (see Table \ref{tab:measurements}) illuminated with an approx. point source from below. From left to right: 3D print with $\A=0.201, 0.26, 0.6, 0.786$ ($\A=0.6$ corresponds to measured \A of the wax). Figure reproduced from Brunton \etal~\protect\shortcite{BruntonArikanTanksaleUrban2018}. \revision{Note that for such extreme illuminating conditions a deviation of \A from perceptual-uniformity can be observed.}}
\label{fig:wax_ships}
\end{figure*}

Here we show some 3D prints leveraging our \A definition. The prints are generated using a recently proposed joint color and translucency multimaterial 3D printing pipeline~\cite{BruntonArikanTanksaleUrban2018}. Figure \ref{fig:lucy_materials} shows the St. Lucy model printed using the sRGB and \A values measured from 3 real samples (see Table \ref{tab:measurements}), with linear transitions between them. Figure \ref{fig:HeadOpaqueTranslucent} shows a head model printed with an sRGB texture and two different \A values. Blurring of geometric and texture details increases for the lower $\A$. Figure \ref{fig:wax_ships} shows ships printed with the color values measured for wax (Table \ref{tab:measurements}), and four different \A values, including that measured for wax. Different shape and magnitude of light transport are visible for different \A values. 

\section{Color Adjustment of Reference Material Renderings}
\label{sec:color_adjust}

To make translucency comparisons easier for materials with spectrally non-uniform absorption, we transfer color in image space from the rendering of the original material to the resulting reference material with the following post-process, which does not change the lightness contrast in the non-specular areas and preserves the translucency cues provided by the reference materials according to Motoyoshi~\shortcite{Motoyoshi2010}. Given an input material, its measurements according to the setup in Section \ref{Subseq::Measurements} along with the resulting reference material, and a rendering of an object with the original material:
\begin{itemize}
	\item We render the same object with the same illumination and viewing conditions as the original with the reference material.
	\item Since specular highlights are dominated by Fresnel reflection, which is similar for both the original and reference materials, but also influenced by absorption, we copy and paste specular highlights from the original rendering into the output rendering using a mask.
	\item Since \A is not a measure of lightness (which is included in the scale-invariant RGB) we adjust the median lightness of the reference rendering to match the median lightness of the original rendering for non-specular pixels. The median lightness is used to mitigate potential influence of remaining specularities.
	\item Finally, for non-specular pixels, we copy the CIELAB a* and b* values from the original rendering to the reference rendering so that they have the same hue.
\end{itemize}
We applied this process to Figures \ref{Figure::Dragon}(c, f), \ref{Figure::Lucy}(e, f, h) and \ref{Figure::Temple}(e, f, h), to make them easier to compare to Figures \ref{Figure::Dragon}(a), \ref{Figure::Lucy}(a) and \ref{Figure::Temple}(a).

\end{document}